\documentclass[a4paper,UKenglish]{lipics_arxiv}

\usepackage{microtype}
\usepackage{amsthm,mathtools}
\usepackage{amssymb}

\newcommand{\E}{{\cal E}}

\newcommand{\chain}{{\cal C}}
\newcommand{\simpl}{{\cal S}}

\newcommand{\eps}{\varepsilon}

\graphicspath{{./figures/}}

\nolinenumbers

\title{Progressive Simplification of Polygonal Curves}

\author{Kevin Buchin}{TU Eindhoven, {Eindhoven, The Netherlands}}{k.a.buchin@tue.nl}{0000-0002-3022-7877}{supported by the Netherlands Organisation for Scientific Research (NWO) under project no.~612.001.207}

\author{Maximilian Konzack$^1$}{TU Eindhoven, {Eindhoven, The Netherlands}}{maximilian.konzack@idiv.de}{}{}

\author{Wim Reddingius}{TU Eindhoven, {Eindhoven, The Netherlands}}{wimreddingius@gmail.com}{}{}

\authorrunning{K.\,Buchin, M.\,Konzack and W.\,Reddingius}

\Copyright{Kevin Buchin, Maximilian Konzack, Wim Reddingius}


\ArticleNo{}

\begin{document}

\maketitle

\begin{abstract}
Simplifying polygonal curves at different levels of detail is an important problem with many applications. Existing geometric optimization algorithms are only capable of minimizing the complexity of a simplified curve for a single level of detail. We present an $O(n^3m)$-time algorithm that takes a polygonal curve of $n$ vertices and produces a set of consistent simplifications for $m$ scales while minimizing the cumulative simplification complexity. This algorithm is compatible with distance measures such as the Hausdorff, the Fr\'echet and area-based distances, and enables simplification for continuous scaling in $O(n^5)$ time. To speed up this algorithm in practice, we present new techniques for constructing and representing so-called shortcut graphs. Experimental evaluation of these techniques on trajectory data reveals a significant improvement of using shortcut graphs for progressive and non-progressive curve simplification, both in terms of running time and memory usage.
 \end{abstract}

\section{Introduction}\label{sec:introduction}
\begin{figure}[tb!]
\centering
\includegraphics[width=\linewidth]{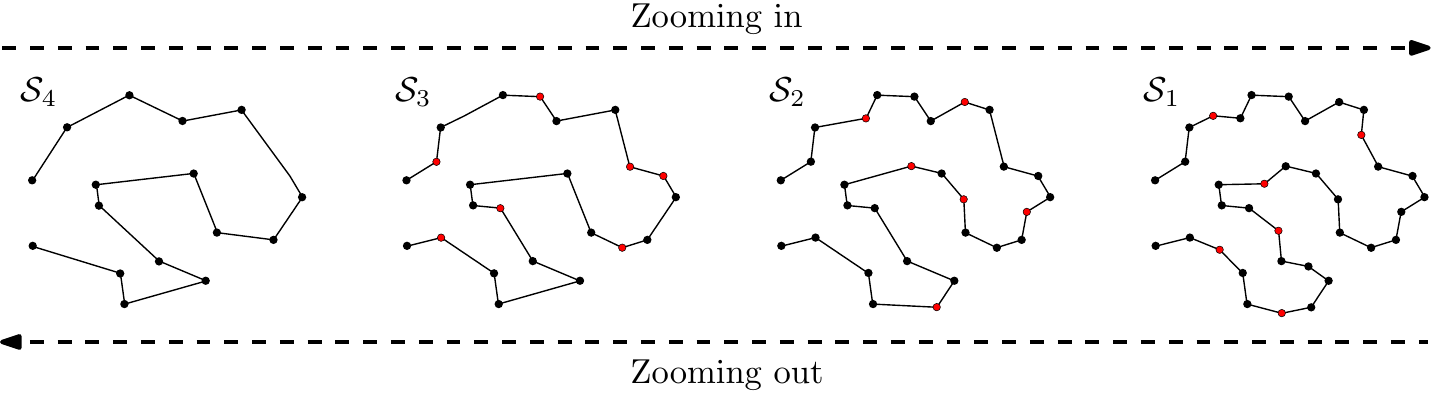}
\caption{A progressive curve simplification for four different levels of detail. Note that when zooming in, we add new vertices (in red) while retaining existing vertices (in black).}
\label{fig:hierarchy}
\end{figure}

Given a polygonal curve as input, the curve simplification problem asks for a polygonal curve that approximates the input well using as few vertices as possible. Because of the importance of data reduction, curve simplification has a wide range of applications. Cartography is such an application, where the visual representation of line features like rivers, roads, and region boundaries needs to be reduced. Most maps nowadays are interactive and incorporate zooming, which requires curve simplification that facilitates different levels of detail. A naive approach would be to simplify each zoom level independently. This, however, has the drawback that the simplifications are not consistent between different scales, resulting in unnecessary flickering when zooming. Therefore, we require \emph{progressive simplification}, that is, a series of simplifications for which the level of detail is progressively increased for higher zoom-levels. This is shown in Figure~\ref{fig:hierarchy}.

Progressive simplifications are used in cartography~\cite{qingsheng2002}.
Existing algorithms for progressive simplification (e.g.~\cite{cao2006spatio}) work by simplifying the input curve, then simplifying this simplification, and so on. More concretely, a common approach is to iteratively discard vertices, such that we always discard the vertex whose removal introduces the smallest error (according to some criterion). For example, the algorithm by Visvalingam~and~Whyatt~\cite{visvalingam1993line} always removes the vertex which together with its neighboring vertices forms a triangle with the smallest area.

Such approaches stand in stark contrast to (non-progressive) curve simplification algorithms that aim to minimize the complexity of the simplification while guaranteeing a (global) bound on the error introduced by the simplification. The most prominent algorithm with a preset error bound was proposed by Douglas and Peucker~\cite{douglas1973algorithms}. However, while heuristically aiming at a simplification with few vertices, this algorithm does not actually minimize the number of vertices. A general technique for the problem of minimizing the number of vertices was introduced by Imai~and~Iri~\cite{Imai88}. Their approach uses \emph{shortcut graphs}, which we describe in more detail below. An efficient algorithm to compute shortcut graphs for the Hausdorff distance was presented by Chan and Chin~\cite{chan1996approximation}. Inspired by the work of Visvalingam~and~Whyatt, Daneshpajouh~et~al.~\cite{daneshpajouh2012computing} defined an error measure for non-progressive simplification by measuring the sum or the difference in area between a simplification and the input curve. Other simplification algorithms minimize the number of vertices while preserving distances~\cite{gudmundsson2007distance} or areas~\cite{bose2006area}.
In the line of these algorithms, the goal of our work is to develop algorithms that solve progressive simplification as an optimization problem.

We assume that a polygonal curve $\chain$ is given as sequence of its vertices, denoted by $\chain = \langle p_1, \ldots, p_n \rangle$.
A (vertex-restricted) \emph{simplification} $\simpl$ of a polygonal curve $\chain$ is an ordered subsequence of $\chain$ (denoted by $\simpl \sqsubseteq \chain$) that includes the first and the last vertex of $\chain$. An \emph{$\eps$-simplification} $\simpl$ is a simplification that ensures that each edge of $\simpl$ has a distance of at most $\eps$ to its corresponding subcurve, where the distance measure can for instance be the Hausdorff or the Fr\'echet distance~\cite{alt1995frechet}. We refer to $\eps$ as \emph{error tolerance} or simply \emph{error} for $\simpl$.
For an ordered pair of vertices $(p_i, p_j)$ of $\chain$, we denote the distance between the segment $(p_i, p_j)$ and the corresponding subcurve by $\eps(p_i, p_j)$. We denote by $(p_i, p_j) \in \simpl$ that $(p_i, p_j)$
is an edge of $\simpl$.

We next define the \emph{progressive simplification problem}. 
Given a polygonal curve $\chain = \langle p_1, \ldots, p_n \rangle$ in $\mathbb{R}^2$ and a sequence $\E = \langle \eps_1, \ldots, \eps_m \rangle$ with $\eps_i \in \mathbb{R}_{>0}$ where $0 < \eps_1 < \ldots < \eps_m$, we want to compute a sequence of 
simplifications $\simpl_1, \simpl_2, \dots, \simpl_m$ of $\chain$ such that

\begin{enumerate}
    \item $\simpl_m \sqsubseteq \simpl_{m-1} \sqsubseteq \ldots \sqsubseteq \simpl_1 \sqsubseteq \chain$ (\emph{monotonicity}),
    \item $\simpl_k$ is an $\eps_k$-simplification of $\chain$,
    \item $\sum_{k=1}^{m} |\simpl_k|$ is minimal.
\end{enumerate}

We refer to a sequence of simplifications fulfilling the first two conditions as \emph{progressive simplification}.
A sequence fulfilling all  conditions is called a \emph{minimal progressive simplification}, and the problem of computing such a sequence is called the \emph{progressive simplification problem}.
 We present an $O(n^3m)$-time algorithm for the progressive simplification problem in the plane.

The cornerstone of progressive simplification is that we require
monotonicity.
This guarantees that, when ``zooming out'',
vertices are only removed and cannot (re)appear.
As error measure, we will
mostly use the Hausdorff distance. This is not essential to the core algorithm,
and we will discuss how to use the Fr\'echet distance \cite{alt1995frechet} or
area-based measures \cite{daneshpajouh2012computing} without affecting the worst-case running time.
Furthermore, our algorithm generalizes to the 
\emph{continuous} version of the problem, wherein $\simpl_{\eps}$
needs to be computed for all $0 \leq \eps < \eps_{M}$.
As in the discrete setting, we require $\simpl_{\eps'} \sqsubseteq
\simpl_{\eps}$ for $\eps'>\eps$; the resulting algorithm minimizes $\int_0^{\eps_{M}} |\simpl_{\eps}|\,d\eps$ in $O(n^5)$ time.
Note that
$\eps_{M}$ is the error tolerance at which we can simplify the curve by the single
line segment $(p_1,p_n)$; thus, we have $\eps_M = \eps(p_1,
p_n)$.

\begin{figure}[t]
 \centering
 \includegraphics[width=.86\linewidth]{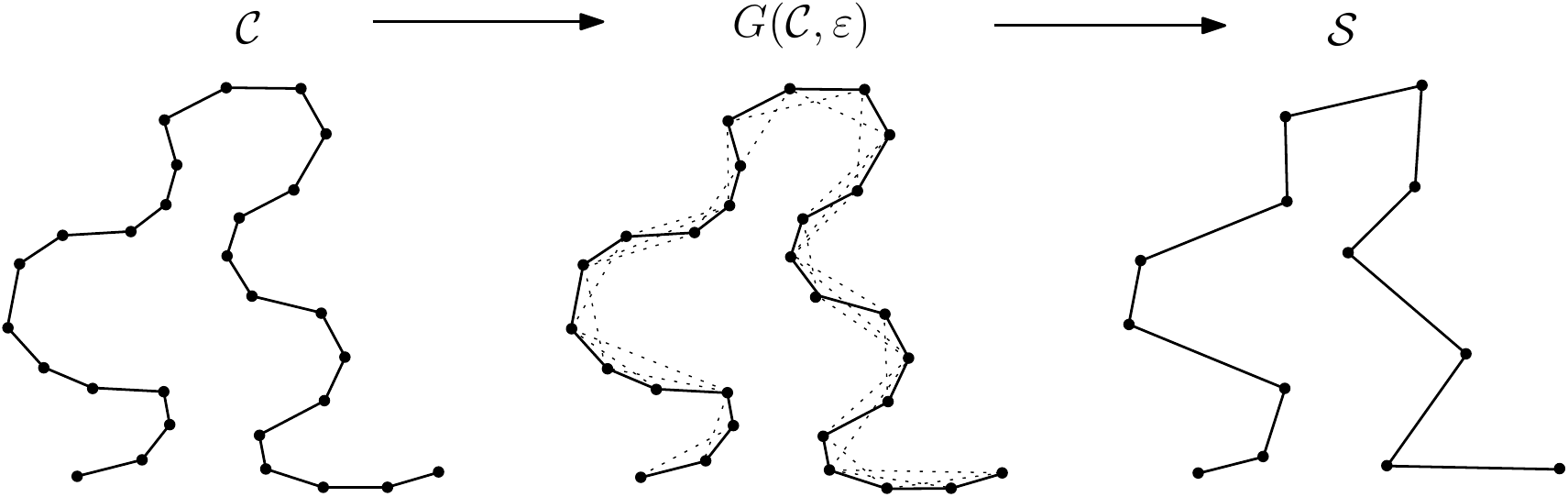}
 \caption{Using shortcut graph $G(\chain, \eps)$ to obtain a simplification $\simpl$ for polygonal curve $\chain$.}
  \label{fig:framework}
\end{figure}

In our algorithms, we make use of the \emph{shortcut graph} as introduced by
Imai~and~Iri~\cite{Imai88}. For a given curve $\chain$, a \emph{shortcut} $(p_i, p_j)$ is an ordered pair ($i < j$) of vertices. Given an error $\eps > 0$, a shortcut $(p_i, p_j)$ is \emph{valid} if $\eps(p_i, p_j) \leq \eps$.
The \emph{shortcut graph} $G(\chain,\eps)$~\cite{Imai88}
represents all valid shortcuts $(p_i, p_j)$ with $1 \le i <
j \le n$. A minimum-link path in this graph, corresponds to a minimal simplification (in the case $m=1$), see Figure~\ref{fig:framework} for an example.

A crucial bottleneck in our algorithms but also in existing algorithms for the (non-progressive, i.e. $m=1$) simplification problem is the construction and space usage of these graphs. We therefore introduce new techniques for the computation and representation of shortcut graphs. These techniques apply to both the progressive and non-progressive simplification problem.

Firstly, we present an algorithm for constructing shortcut graphs for many levels of detail efficiently. To date, it has been known only how to compute shortcut graphs under the Hausdorff distance in subcubic time if the error $\eps$ is given up front. Our algorithm computes the errors $\eps(p_i,p_j)$ of all shortcuts $(p_i, p_j)$ by incrementally constructing an augmented convex hull of contiguous subsequences of the curve.

Secondly, we introduce a compressed representation of the shortcut graph that employs so-called \emph{shortcut intervals}. We show that this representation has linear space complexity in practice, and illustrate how we can use this to compute shortest paths in only $O(n \log n)$ time in practice, instead of using breadth-first search in $O(n^2)$ time.

In our experiments, we compare our minimal progressive simplification algorithm with several natural heuristics. Furthermore, we evaluate our constructing of the shortcut graph for many levels and our compressed representation of shortcut graphs.

\section{Optimal Progressive Simplification}\label{sec:progr_simplification}
Many applications, such as cartography or GIS, require computing or visualizing curve features on many spatial scales.
We call a series of simplifications that is consistent for varying spatial scales progressive simplification.
In this section, we first show how to solve the progressive simplification problem  
in $O(n^3m)$ time for $m$ scales. We then generalize our algorithm to solve the continuous progressive simplification problem in $O(n^5)$ time.

\subsection{Progressive simplification for \emph{m} scales}
By the monotonicity property of the progressive simplification problem (see Condition 1
in the definition in Section~\ref{sec:introduction}), we require that all
vertices within a simplification $\simpl_k$ of the sequence must also occur
within all subsequent simplifications $\simpl_l$ with $k < l$.
Adding shortcuts to a specific simplification thus influences the structure of the other simplifications.
Inspired by the shortcut graph representation by Imai and Iri \cite{Imai88},
wherein each valid shortcut has unit costs, we decided to model each shortcut
$(p_i, p_j)$ in the shortcut graph $G(\chain, \eps_k)$ at scale $\eps_k$ by a cost value \mbox{$c_{i,j}^k \in \mathbb{N}$},
describing the cost of including $(p_i, p_j)$ in $\simpl_k$.
We use
the Hausdorff distance as an error measure to determine whether a shortcut is
valid, but since the shortcut graph is flexible to use any error measures, we
can employ any other distance measure for our algorithms. In particular for the
Fr\'echet distance~\cite{alt1995frechet} and area-based
distances~\cite{daneshpajouh2012computing}, we can use easily compute
whether a shortcut is valid in $O(n)$ time, and therefore use these measures
without changing the worst-case running time.  We obtain a cost value
$c_{i,j}^k$ for a shortcut $(p_i, p_j) \in G(\chain, \eps_k)$ by minimizing
the costs of all possible shortcuts in $\langle p_i, \ldots, p_j
\rangle$ at lower scales recursively. The dynamic program is defined as follows:
\begin{align*}
c_{i,j}^k = \begin{cases}
  1 & \text{if}\ k = 1\\
  1 +  {\displaystyle\min_{\pi \in \prod^{k-1}_{i,j}} \sum_{(p_x, p_y) \in \pi} c_{x,y}^{k - 1}} & \text{if}\  1 < k \leq m
\end{cases}
\end{align*}


where $\prod^k_{i,j}$ denotes the set of all paths in $G(\chain, \eps_k)$ from $p_i$ to $p_j$.

We construct the sequence of simplifications from $\simpl_m$ down to
$\simpl_1$. First, we compute $\simpl_m$ by returning the shortest path from
$p_1$ to $p_n$ in $G(\chain, \eps_m)$ using the computed cost values at scale
$m$. Next, we compute a shortest path $P$ from $p_i$ to $p_j$ in $G(\chain,
\eps_{m - 1})$ for all shortcuts $(p_i, p_j) \in \simpl_m$. Simplification
$\simpl_{m-1}$ is then constructed by linking these paths $P$ with each other.
We build all other simplifications in this manner until $\simpl_1$ is constructed.

The algorithm starts with constructing the shortcut graphs $G(\chain, \eps_m), \ldots, G(\chain, \eps_1)$. For most distance measures, the distance of shortcut $(p_i, p_j)$ to the subcurve $\langle p_i, \ldots, p_j \rangle$ can be determined in $O(j - i)$ time. For such measures, constructing these graphs naively takes $O(n^3m)$ time because we need to determine for each shortcut ($O(n^2)$ shortcuts) which of the $O(n)$ vertices of the curve are approximated by the shortcut at each spatial scale ($m$ scales). By employing the algorithm by Chan~and~Chin~\cite{chan1996approximation}, we can compute it in $O(n^2m)$ time for the Hausdorff distance.


We compute all cost values from scale $k = m$ up to $1$ by assigning a weight $c_{i,j}^k$ to each shortcut $(p_i, p_j) \in G(\chain, \eps_k)$. For each shortcut $(p_i, p_j) \in G(\chain, \eps_k)$, we compute $c_{i,j}^k$ by finding a shortest path $\pi$ in $G(\chain, \eps_{k - 1})$ from $p_i$ to $p_j$, minimizing $\sum_{(p_x, p_y) \in \pi} c_{x,y}^{k - 1}$. This shortest path computation resembles the one from Imai and Iri \cite{Imai88} for the original $\min-\#$ simplification problem.

We can employ any shortest path algorithm for this, such as Dijkstra's algorithm.
On each scale $k$, we need to run Dijkstra's algorithm on $O(n)$ source nodes of $G(\chain, \eps_k)$. This yields a worst case running time of $O(n^3m)$,
because the shortcut graph can have $O(n^2)$ edges in the worst case and we need to run Dijkstra's algorithm on each source node for each spatial scale.

We increment
$c_{i,j}^k = c_{i,j}^{k - 1} + 1$ for any shortcut $(p_i, p_j) \in G(\chain, \eps_{k - 1})$.
By doing so, we
avoid recomputations of shortest paths and reuse cost values whenever
necessary.


If $(p_i, p_j)$ is a valid shortcut in $G(\chain, \varepsilon_{k-1})$ for any $1 < k \le m$, then it follows that $c_{i,j}^k = c_{i,j}^{k - 1} + 1$.
We prove the following lemma:
\begin{lemma}\label{lemma:hierarchical_weights_increment}
	For any $1 < k \leq m$ and $(p_i, p_j) \in G(\chain, \eps_{k - 1})$, it follows that $c_{i,j}^k = c_{i,j}^{k - 1} + 1$.
\end{lemma}
\begin{figure*}[b!]
	\centering
	\includegraphics[width=\textwidth]{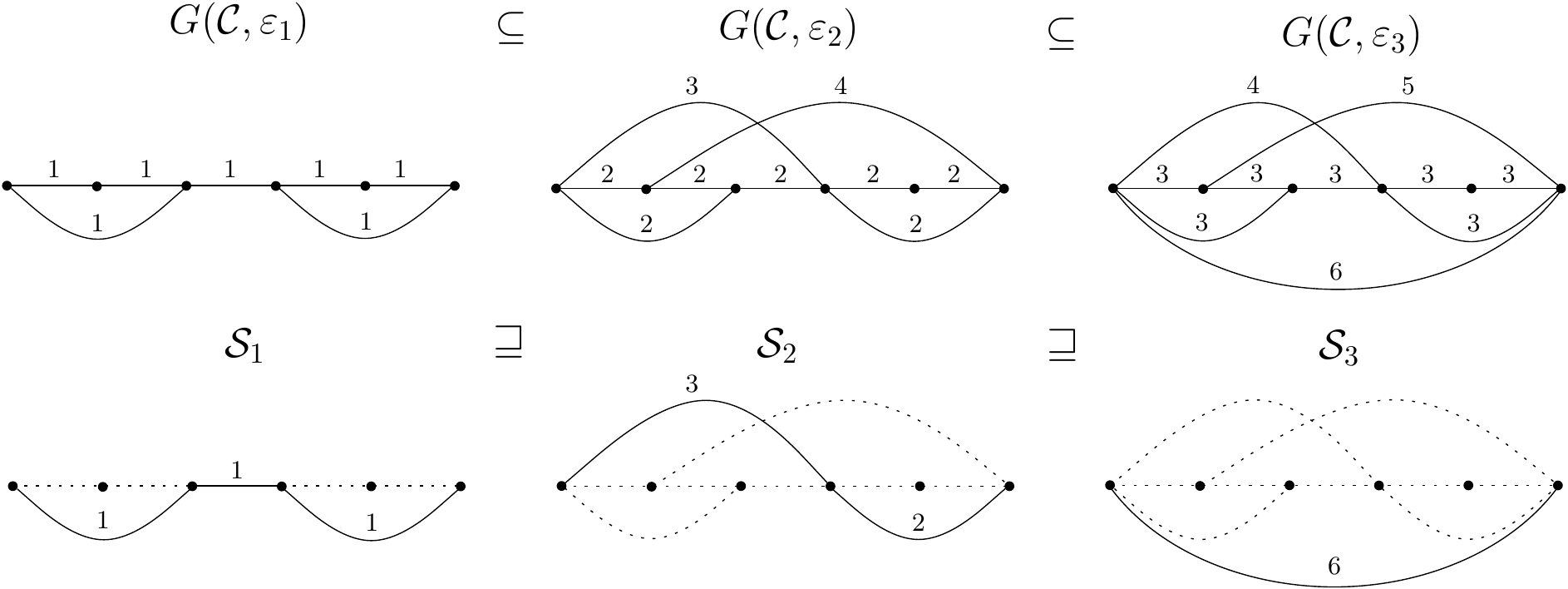}
	\caption{Using shortcut graphs weighted by cost to find a minimal progressive simplification.}
	\label{fig:hierarchical_weights}
\end{figure*}
\begin{proof}[Proof of Lemma~\ref{lemma:hierarchical_weights_increment}]
	Assume $(p_i, p_j) \in G(\chain, \eps_{k - 1})$. Because $\eps_{k} > \eps_{k - 1}$, we have $G(\chain, \eps_{k - 1}) \subseteq G(\chain, \eps_{k})$, which implies that $(p_i, p_j) \in G(\chain, \eps_{k})$. See Figure~\ref{fig:hierarchical_weights} for an example showing the graphs $G(\chain, \eps_{k})$ (and their use in the algorithm).
	Now let us make the following case distinction:\\


	\noindent \textbf{Case $k = 2$:} Because $(p_i, p_j) \in G(\chain, \eps_1)$, we can derive $c_{i,j}^{2} = c_{i,j}^{1} + 1$ as follows:
	\begin{align*}
	c_{i,j}^{2}\hspace{3pt} &=\hspace{5pt} 1 + \min_{\pi \in \prod^{1}_{i,j}} \sum_{(p_x, p_y) \in \pi} c_{x,y}^1 =\hspace{5pt} 1 + \min_{\pi \in \prod^{1}_{i,j}} \sum_{(p_x, p_y) \in \pi} 1
	=\hspace{5pt} 1 + 1
	=\hspace{5pt} c_{i,j}^{1} + 1
	\end{align*}



	\noindent \textbf{Case $k > 2$:} We prove $c_{i,j}^{k} = c_{i,j}^{k - 1} + 1$ by showing that both the upper and lower bound of $c_{i,j}^{k}$ are equal to $c_{i,j}^{k - 1} + 1$. Because $(p_i, p_j) \in G(\chain, \eps_{k - 1})$, we can derive the upper bound as follows:


	\begin{align*}
	c_{i,j}^{k}\hspace{3pt} &=\hspace{5pt} \min_{\pi \in \prod^{k - 1}_{i,j}} \sum_{(p_x, p_y) \in \pi} c_{x,y}^{k - 1} + 1
	\leq\hspace{5pt} \sum_{(p_x, p_y) \in \langle p_i, p_j \rangle} c_{x,y}^{k - 1} + 1
	=\hspace{5pt} c_{i,j}^{k - 1} + 1
	\end{align*}

	\noindent Now let us prove the lower bound.
	\begin{align*}
	c_{i,j}^{k}\hspace{3pt} &=\hspace{5pt} 1 + \min_{\pi \in \prod^{k - 1}_{i,j}} \sum_{(p_x, p_y) \in \pi} c_{x,y}^{k - 1}  =\hspace{5pt} 1 + \min_{\pi \in \prod^{k - 1}_{i,j}} \sum_{(p_x, p_y) \in \pi} (1 + \min_{\pi' \in \prod^{k - 2}_{x,y}} \sum_{(p_a, p_b) \in \pi'} c_{a,b}^{k - 2}) \\
	&\geq\hspace{5pt} 1 + 1 + \min_{\pi \in \prod^{k - 1}_{i,j}} \sum_{(p_x, p_y) \in \pi} \min_{\pi' \in \prod^{k - 2}_{x,y}} \sum_{(p_a, p_b) \in \pi'} c_{a,b}^{k - 2}  \geq\hspace{5pt} 1 + 1 + \min_{\pi \in \prod^{k - 2}_{i,j}} \sum_{(p_x, p_y) \in \pi} c_{x,y}^{k - 2} \\
	&=\hspace{5pt} c_{i,j}^{k - 1} + 1
	\end{align*}


	\noindent The last inequality holds, since a path from $p_i$ to $p_j$ constructed by concatenating several shortest subpaths is always at least as long as a single shortest path from $p_i$ to $p_j$.
\end{proof}

We prove that our simplification algorithm returns a valid and minimal solution for the progressive simplification problem.
Let $\langle \simpl_1, \ldots, \simpl_m \rangle$ be a sequence of simplifications computed by our algorithm. By constructing the simplifications from scale $m$ down to $1$, it follows that for any shortcut $(p_i, p_j) \in \simpl_{k}$ with $1 < k \leq m$, there exists a subsequence $\langle p_i, \ldots, p_j \rangle \sqsubseteq \simpl_{k - 1}$, and thus $\simpl_{k} \sqsubseteq \simpl_{k - 1}$. Furthermore, each simplification $\simpl_k$ has a maximum Hausdorff distance $\eps_k$ to $\chain$ since it contains only edges from $G(\chain, \eps_k)$.

It remains to show that we minimize $\sum_{i=1}^m |\simpl_i|$. We therefore
define a set of shortcuts $\simpl_k^{i,j}$ for any $ 1 \leq i < j \le n$ and $1\leq k \leq m$ as
$\simpl_k^{i,j} = \{ \ (p_x, p_y) \in \simpl_k \ | \ x \leq i < j \leq y \ \}$.

Thus, $\simpl_k^{i,j}$ includes all line segments of $\simpl_k$ that span the
subcurve $\langle p_i, \ldots, p_j \rangle$ with an error of at most
$\varepsilon_k$ to $\chain$. $|S_k^{i,j}|$ then is the number of shortcuts in
simplification $\simpl_k$ covering $(p_i, p_j)$.

\begin{lemma}\label{lemma:hierarchical_weight_meaning}
If the line segment $(p_i, p_j)$ is part of simplification $\simpl_k$, then the
associated cost value $c_{i,j}^k = \sum_{\ell = 1}^k |\simpl_\ell^{i,j}|$ for any $1 \leq k \leq m$ and $1 \leq i < j \leq n$.
\end{lemma}
\begin{proof}
We show $c_{i,j}^k = \sum_{\ell = 1}^k |\simpl_\ell^{i,j}|$ by induction on $k$ using the following induction hypothesis:
For any $(p_x, p_y) \in \simpl_k$, it holds that $c_{x,y}^k = \sum_{\ell = 1}^k |\simpl_\ell^{x,y}|$ (IH).

\noindent \textbf{Base $k = 1$:} Take any shortcut $(p_i, p_j) \in \simpl_1$. It follows that $\simpl_1^{i,j} = \{ (p_i, p_j) \}$, and therefore $|\simpl_1^{i,j}| = 1$. We deduce that $
c_{i,j}^{1} = 1 = \sum\nolimits_{\ell = 1}^k 1 = \sum\nolimits_{\ell = 1}^k |\simpl_1^{i,j}|$.

\noindent \textbf{Step $k > 1$:}
Take any line segment $(p_i, p_j) \in
\simpl_{k + 1}$. Thus, we observe $(p_i, p_j) \in G(\chain, \eps_{k + 1})$, $\simpl_{k + 1}^{i,j} = \{ (p_i, p_j) \}$, and $|\simpl_{k + 1}^{i,j}| = 1$.

Consider any $1\leq \ell \leq k$ and a path $\pi \in \prod^k(p_i, p_j)$ such that $\sum\nolimits_{(p_x, p_y) \in \pi} |\simpl^{x,y}_\ell|$ is minimal. We now derive
that $\pi = \simpl^{i,j}_\ell$ such that $\simpl^{x,y}_\ell$ is minimal for all $(p_x, p_y) \in \pi$. Note that $\pi = \simpl^{i,j}_\ell \subseteq G(\chain, \eps_\ell) \subseteq G(\chain, \eps_k)$ since $\eps_k \geq \eps_\ell$. We observe that $\pi$ is both in $\prod^\ell(p_i, p_j)$ and $\prod^k(p_i, p_j)$. It thus follows that:
\begin{align}\label{eq:changing_scale}
\min_{\pi \in \prod^{k}_{i,j}} \sum_{(p_x, p_y) \in \pi}  |\simpl_\ell^{x,y}| = \min_{\pi \in \prod^{\ell}_{i,j}} \sum_{(p_x, p_y) \in \pi}  |\simpl_\ell^{x,y}|
\end{align}

From $\pi = \simpl^{i,j}_\ell$ it follows that $\simpl^{x,y}_\ell \cap \simpl^{y,z}_\ell = \emptyset$ for any $(p_x, p_y)$ and $(p_y, p_z)$ in  $\pi$.
Combining $\simpl^{x,y}_\ell$ for all $(p_x, p_y) \in \pi$ yields a non-overlapping sequence of shortcuts from $p_i$ to $p_j$. This gives us:
\begin{align}\label{eq:combining}
|\simpl^{i,j}_\ell| = \min_{\pi \in \prod^\ell_{i, j}} \sum_{(p_x, p_y) \in \pi} |\simpl^{x,y}_\ell|
\end{align}

\noindent
We now derive the following:
\begin{align*}
c_{i,j}^{k + 1} &\stackrel{\text{(IH)}}{=} 1 + \min_{\pi \in \prod^{k}_{i,j}} \sum_{(p_x, p_y) \in \pi} \sum_{\ell = 1}^k |\simpl_\ell^{x,y}|
\stackrel{\eqref{eq:changing_scale}}{=} 1 + \sum_{\ell = 1}^k \min_{\pi \in \prod^{\ell}_{i,j}} \sum_{(p_x, p_y) \in \pi}  |\simpl_\ell^{x,y}|
\stackrel{\eqref{eq:combining}}{=} 1 + \sum_{\ell = 1}^k |\simpl_\ell^{i,j}| \\
&\stackrel{|\simpl_{k + 1}^{i,j}| = \{(p_i, p_j)\}}{=} \sum_{\ell = 1}^{k + 1} |\simpl_\ell^{i,j}|
\end{align*}\end{proof}

By observing the combined size of the computed simplification is minimal, we obtain the following theorem.
\begin{theorem}\label{theorem:hierarchical_opt}
Given a polygonal curve with $n$ vertices in the plane, and $0 \leq \varepsilon_1 < \ldots < \varepsilon_m$, a minimal progressive simplification can be computed in $O(n^3 m)$ time under distance measures for which the validity of a shortcut can be computed in $O(n)$ time. This includes the Fr\'echet, the Hausdorff and area-based measures.
\end{theorem}

\begin{proof}[Proof of Theorem~\ref{theorem:hierarchical_opt}]
	It remains to prove that the combined size of the simplifications computed by our algorithm is minimal. Let  $\langle \simpl'_1, \ldots, \simpl'_m \rangle$ be a sequence of simplifications of a minimal progressive simplification, and let $\langle \simpl_1, \ldots, \simpl_m \rangle$ be the sequence computed by our algorithm.

	Let us derive the following:
	\[
	\min_{\pi \in \prod^m_{1,n}} \sum_{(p_x, p_y) \in \pi} c_{x,y}^m \stackrel{\eqref{lemma:hierarchical_weight_meaning}}{=}\min_{\pi \in \prod^m_{1,n}} \sum_{(p_x, p_y) \in \pi} \sum_{k = 1}^m |\simpl^{x,y}_k|
	\stackrel{\eqref{eq:changing_scale}}{=}\sum_{k = 1}^m \min_{\pi \in \prod^\ell_{1,n}}  \sum_{(p_x, p_y) \in \pi} |\simpl^{x,y}_k|
	\stackrel{\eqref{eq:combining}}{=}\sum_{k = 1}^m |\simpl_k|
	\]

	Hence, the algorithm produces a simplification that minimizes the cumulative cost of shortcuts in $\simpl_m$. Because $\simpl_{i+1} \sqsubseteq \simpl_{i}$; the algorithm produces a set of simplifications in which each simplification consists of edges from the corresponding shortcut graph such that the cumulative number of vertices is minimized.

	We further know that any minimal simplification $\simpl_k'$ is a path in $G(\chain, \eps_k)$ since it strictly connects shortcuts with an error of at most $\eps_k$.

	We conclude that $\sum_{k = 1}^m |\simpl_k| \leq \sum_{k = 1}^m |\simpl_k'|$.
\end{proof}

\subsection{Continuous Simplification}\label{sec:continuous}
Before solving the continuous case, we consider
\emph{weighted} progressive simplification, wherein the objective is to minimize $\sum_{k = 1}^m w_k |\simpl_k|$ (with $w_k \ge 0$), thus
the weighted cumulative size of the simplifications.
For the weighted progressive simplification, we use the following cost function for each shortcut $(p_i, p_j) \in G(\chain, \eps_k)$:
if $k = 1$, $c_{i,j}^k = w_1$ else $c_{i,j}^k = w_k + \min_{\pi \in \prod^{k-1}_{i,j}} \sum_{(p_x, p_y) \in \pi} c_{x,y}^{k - 1}$.
The
proofs for the regular/unweighted case trivially extended to apply to this updated cost function.
The main reason to consider the weighted case is that it helps us solving the continuous
progressive simplification problem.

\begin{theorem}\label{theorem:continuous}
Given a polygonal curve with $n$ vertices in the plane, a minimal continuous progressive simplification can be computed in $O(n^5)$ time under distance measures for which the validity of a shortcut can be computed in $O(n)$ time. This includes the Fr\'echet, the Hausdorff and area-based measures.
\end{theorem}
\begin{proof}
    Consider the error tolerances $\varepsilon(p_i, p_j)$ of all possible line
    segments $(p_i, p_j)$ with $i < j$ with respect to the Hausdorff distance
    (or another distance measure). Then, let $\mathcal{E} := \langle \varepsilon_1, \dots,
    \varepsilon_{\binom{n}{2}} \rangle$ be the sorted sequence of these error tolerances
    based on their value.
    Let $M$
    be the index of the corresponding $\varepsilon_M$ in this sorted sequence
    $\mathcal{E}$ for the line segment $(p_1, p_n)$; thus $\varepsilon_M = \varepsilon(p_1, p_n)$.
Note that it is possible that $M < \binom{n}{2}$, but there is no reason to use
any $\varepsilon > \varepsilon_{M}$, since at this point we already have
simplified the curve to a single line segment, $(p_1, p_n)$.

In a minimal-size progressive simplification it holds that $\simpl_{\varepsilon} = \simpl_{\varepsilon_i}$ for all $\varepsilon \in [\varepsilon_i, \varepsilon_{i+1})$. This can be shown by contradiction: if  $\simpl_\varepsilon$ would be smaller, we could decrease the overall size by setting all $\simpl_{\varepsilon'}$ with $\varepsilon' \in [\varepsilon_i, \varepsilon]$ to $\simpl_{\varepsilon}$. Therefore, in a minimal continuous progressive simplification we have
$
\int_0^{\varepsilon_M} |\simpl_{\varepsilon}|\,d\varepsilon = \sum_{k=1}^{M-1} (\varepsilon_{k+1} - \varepsilon_k )|\simpl_{\varepsilon_k}|$.
Thus, we can solve the continuous progressive simplification problem by reducing it to the weighted progressive simplification problem with $O(n^2)$ values $\varepsilon_k$ and weights $w_k = \varepsilon_{k+1} - \varepsilon_k$.
\end{proof}

\section{Constructing the Shortcut Graph for Arbitrary Scale}\label{sec:convex_hulls}
As described in Section \ref{sec:progr_simplification}, the first step in producing a minimal progressive simplification is to construct shortcut graphs for $m$ different error tolerances. One approach is to construct each graph $G(\chain, \eps_k)$ independently; deciding for every line segment whether or not it is a shortcut for a given error $\eps_k$. However, this is likely to cause overhead when simplifying for many different levels of detail (e.g. Section \ref{sec:continuous}). Thus, instead of independently deciding for every error tolerance whether each shortcut is valid or not, we can determine the error $\eps(p_i,p_j)$ of each shortcut $(p_i, p_j)$, i.e., the ``distance'' from $(p_i, p_j)$ to its subcurve $\langle p_i, \ldots, p_j \rangle$. Afterwards, a shortcut graph can be constructed for any error by simply filtering on these errors. Assuming we use a distance measure which computes the error tolerance of a shortcut $(p_i, p_j)$ in $O(j - i)$ time, we spend only $O(n^3)$ time using this approach. This is an improvement over independently constructing the shortcut graphs in $O(n^3m)$ time.

In this section, we show how we can further improve this bound to $O(n^2 \log n)$ time for the Hausdorff distance. 
The error $\eps(p_i,p_j)$ of shortcut $(p_i, p_j)$ is the distance from the furthest point $X_{i,j}$ in $\langle p_i, \ldots, p_j \rangle$ to line segment $(p_i, p_j)$. Therefore, $X_{i,j}$ must be a point on the convex hull enclosing all points in $\langle p_i, \ldots, p_j \rangle$.

To illustrate our approach, let us consider the simpler problem of searching for the vertex $X'_{i,j}$ furthest from the line through $p_i$ and $p_j$ (instead of furthest from the line segment) for each pair of vertices $p_i$, $p_j$. For every $p_i$ we could incrementally construct a convex hull $CH_i$ of the vertices $p_i, \ldots, p_j$, for $j = i+1, \ldots, n$. We can insert the next $p_j$ in $O(\log n)$ time by storing $CH_i$ suitably, and then find $X'_{i,j}$ by an extreme point query also in $O(\log n)$ time. In this way, we could compute all $X'_{i,j}$ in $O(n^2 \log n)$ time.

However, we want to compute the furthest point $X_{i,j}$ not $X'_{i,j}$. By computing the furthest point from the ray from $p_i$ through $p_j$ and from the ray from $p_j$ through $p_i$, we ensure that one of the points is $X_{i,j}$. To compute the furthest points from these rays, we can use the same approach as for $X'_{i,j}$ but need to suitably augment our convex hull data structure.
In summary, we obtain the following theorem.

\begin{theorem}\label{theorem:convexhull}
Given a polygonal curve $\langle p_1, \ldots, p_n \rangle$ in the plane, we can compute for all $1 \leq i < j \leq n$ the Hausdorff distance between line segment $(p_i,p_j)$ and the subcurve $\langle p_i, \ldots, p_j \rangle$ in $O(n^2 \log n)$ time.
\end{theorem}

We discuss the construction for this theorem now.
To compute $\eps(p_i,p_j)$ for all pairs $p_i,p_j$,
we construct a convex hull $CH_i$ for all vertices $p_i \in \chain$, represented by an upper hull $CH_i^t$ and a lower hull $CH_i^b$. Both $CH_i^t$ and $CH_i^b$ are represented using a balanced binary search tree ordered by the $x$-coordinates of its points. We construct $CH_i^t$ and $CH_i^b$ by incrementally inserting all points from $p_i$ up to $p_n$, such that, after inserting some point $p_j$, $CH_i$ will represent a convex hull enclosing $\langle p_i, \ldots, p_j \rangle$.

To simplify the description of the algorithm, we assume without of generality that $(p_i, p_j)$ is a horizontal line segment with $p_i$ to the left. We subdivide the area around $(p_i, p_j)$ into four regions: $T$(op), $B$(ottom), $L$(eft), and $R$(ight).

\begin{figure}[htb!]
	\centering
	\includegraphics[width=.8\linewidth]{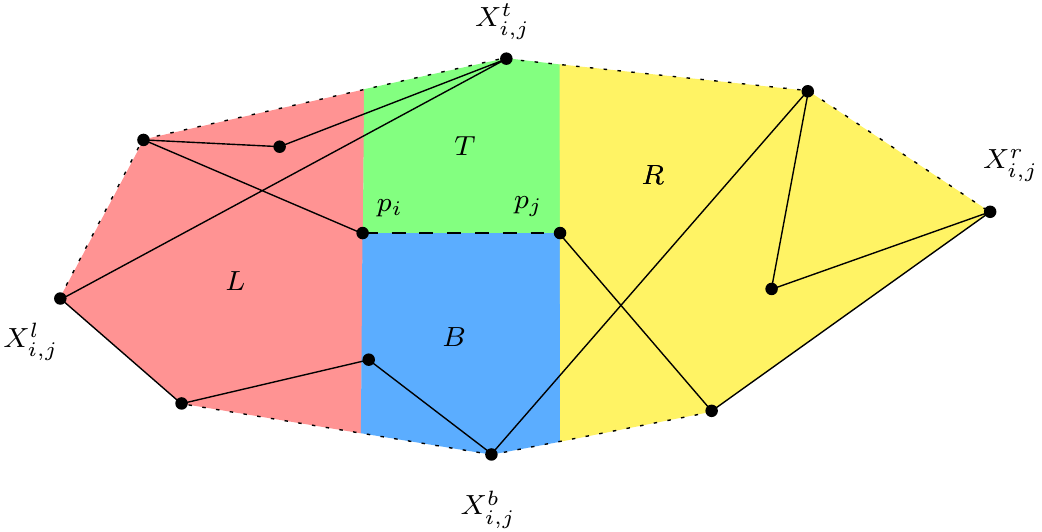}
	\caption{Division of the convex hull into regions $L$ (red), $R$ (yellow) , $T$ (green) and $B$ (blue).}
	\label{fig:convexhull_regions}
\end{figure}

After inserting a point $p_j$ into the convex hull $CH_i$, we first extract candidates for $X_{i,j}$ by finding extreme points on the convex hull in the directions orthogonal to the line segment $(p_i, p_j)$ in $O(\log n)$ time per query. By finding extreme point in the directions orthogonal to $(p_i, p_j)$, we identify the best candidates for $X_{i,j}$ in regions $T$ and $B$, but we may miss points in regions $L$ and $R$. An example of such a scenario is given in Figure~\ref{fig:convexhull_regions}. Therefore, we also need to determine the furthest point $X^{l}_{i,j}$ from $p_i$ in region $L$, and the furthest point $X^{r}_{i,j}$ from $p_j$ in region $R$.

\begin{figure*}[ht!]
	\centering
	\includegraphics[width=\textwidth]{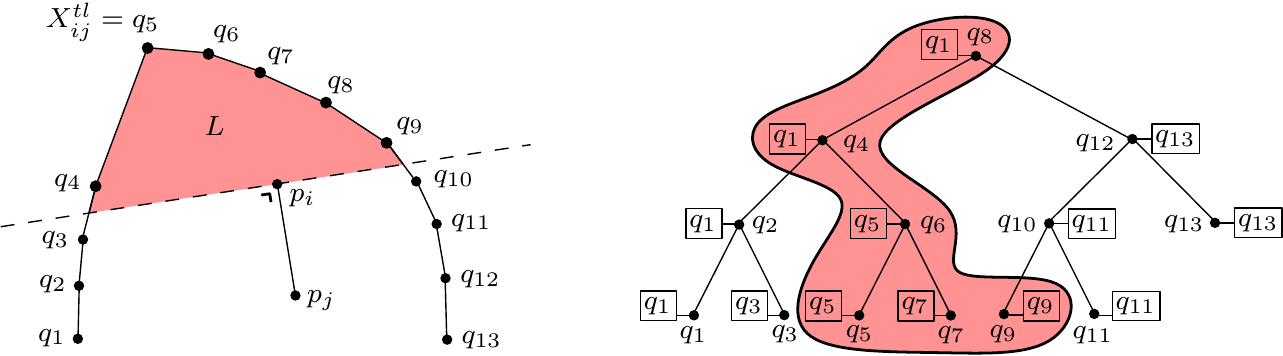}
	\caption{Annotating the binary search tree of the upper convex hull $CH_i^t =
		\langle q_1, \ldots, q_{13} \rangle$ to find $X_{i,j}^{tl}$. Case 1 applies at node $q_6$, meaning we can use its annotation ($q_5$) as a candidate for $X_{i,j}^{tl}$.}
	\label{fig:convexhull_enhanced}
\end{figure*}

We obtain $X^{l}_{i,j}$ by annotating the convex hull and extracting candidates $X^{tl}_{i,j}$ and $X^{bl}_{i,j}$ using a range query on $CH_i^t$ and $CH_i^b$ respectively. For example, we determine $X^{tl}_{i,j}$ by maintaining an annotation of the root node $p_r$ of each subtree $T_r \in CH_i^t$ with the furthest point in $T_r$ to $p_i$. The root node of $CH_i^t$ is therefore annotated with the point in $CH_i^t$ furthest from $p_i$. An example of such a tree annotation is shown in Figure~\ref{fig:convexhull_enhanced}.

Such an annotation allows us to search for subtrees of $CH_i^t$ which lie completely inside $L$, and use the annotation of these subtrees as candidates for $X^{tl}_{i,j}$. We traverse the search tree and check the following for every node $q_c$ rooted at subtree $T_c = \langle q_{min}, \ldots, q_{max} \rangle$:

\begin{enumerate}
	\item If both $q_{min}$ and $q_{max}$ are inside $L$, we know by the convexity of the hull that the entire subtree lies inside $L$. We thus consider the annotation of $q_c$ as a candidate for $X_{tl}^{ij}$.
	\item If either $q_{min}$ or $q_{max}$ are inside $L$, we continue our search by traversing to both children of $q_c$.
	\item Otherwise, stop the search.
\end{enumerate}

\begin{figure}[htb!]
	\centering
	\begin{subfigure}{.42\textwidth}
		\includegraphics[width=\linewidth]{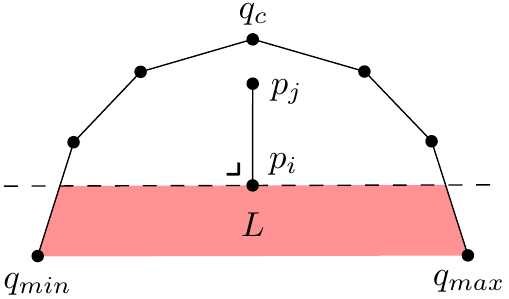}
		\caption{$p_j$ above $p_i$ and $q_{min}$, $q_{max}$ both in $L$}
		\label{fig:convexhull_degenerate2}
	\end{subfigure}
	\qquad \qquad
	\begin{subfigure}{.42\textwidth}
		\includegraphics[width=\linewidth]{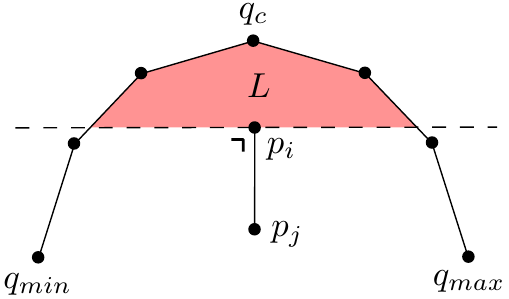}
		\caption{$p_j$ below $p_i$ and $q_{min}$, $q_{max}$ not in $L$}
		\label{fig:convexhull_degenerate1}
	\end{subfigure}
	\caption{Degenerate cases where $p_i$ lies horizontally between $q_{min}$ and $q_{max}$.}
	\label{fig:convexhull_degenerate}
\end{figure}

However, there are two degenerate cases to consider whenever $p_i$ lies horizontally between $q_{min}$ and $q_{max}$. Firstly, $p_j$ might lie above $p_i$ and both $q_{min}$ and $q_{max}$ lie inside $L$, as illustrated in Figure \ref{fig:convexhull_degenerate2}. In this scenario, Case 1 applies, yet not all points in $T_c$ lie inside $L$, meaning we cannot reliably use its annotation. Secondly, we may have the opposite, where $p_j$ lies below $p_i$, and neither $q_{min}$ nor $q_{max}$ lies inside $L$. An example of this is shown in Figure \ref{fig:convexhull_degenerate1}. Case~3 applies, yet there are points from $T_c$ that lie inside $L$, meaning we should further explore descendants of $q_c$. Because in either scenario we have not yet reached a subtree which lies completely inside $L$, we handle these degenerate cases by traversing to both children of $q_c$, like we do in Case 2.

For region $R$, we cannot reuse this approach to determine the furthest point $X^r_{i,j}$, since $p_i$ is the first point added to the convex hull, and $p_j$ the last. We solve this by running this annotation algorithm on the reversed sequence of $\chain$ as well, constructing each convex hull $CH_j$ by incrementally inserting all points from $p_j$ down to $p_1$. For every shortcut $(p_i, p_j)$, we determine $\eps(p_i,p_j)$ by computing $X^t_{i,j}, X^b_{i,j}$ and $X^{l}_{i,j}$ during the forward traversal of $\chain$, and $X^r_{i,j}$ during its reverse traversal.

Overall, we perform $O(n^2)$ insertions and queries, resulting in a total running time of $O(n^2 \log n)$.

\section{Compressing the Shortcut Graph}\label{sec:intervals}
For many types of spatial data, such as movement trajectories or line features on a map, consecutive points $p_i$ and $p_j$ are expected to be spatially close. We can therefore presume that, if $(p_x, p_i)$ is a valid shortcut for some point $p_x$, then $(p_x, p_j)$ is most likely a shortcut as well. We can exploit this phenomenon to represent shortcut graphs using so-called \emph{shortcut intervals}, which are contiguous subsequences of $\chain$ with which a particular point forms shortcuts. Alewijnse~et~al.~\cite{alewijnse2014framework} utilised this fact in a similar approach to speed up trajectory segmentation.

Formally, any shortcut interval for a point $p_i$ and error tolerance $\varepsilon$ is a maximal interval $[x,y]$ where all shortcuts $(p_i, p_j)$ for $x \le j \le y$ are valid for $\eps$. Instead of representing a shortcut graph with a graph $G(\chain, \eps)$ that explicitly stores all edges, we represent the shortcuts using a \emph{shortcut interval set} $I(\chain, \eps) = \langle I_1(\varepsilon), \dots, I_n(\varepsilon) \rangle$, where $I_i(\eps)$ is a sequence of all shortcut intervals for $p_i$ and $\eps$.

\begin{figure*}[htb]
\centering
\begin{subfigure}{.185\textwidth}
    \includegraphics[width=\linewidth]{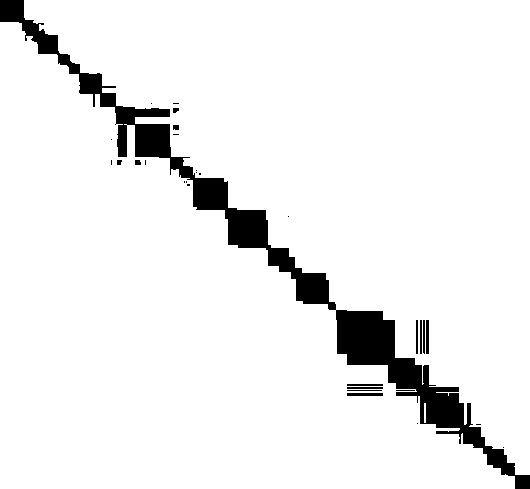}
    \caption{$I(\chain, \eps_1)$}
    \label{fig:matrix1}
\end{subfigure}
\begin{subfigure}{.185\textwidth}
    \includegraphics[width=\linewidth]{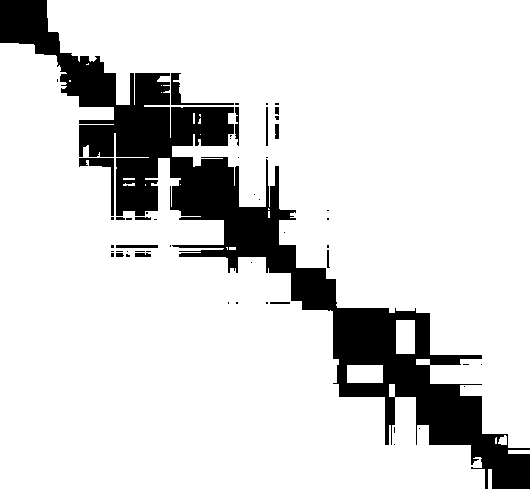}
    \caption{$I(\chain, \eps_2)$}
    \label{fig:matrix2}
\end{subfigure}
\begin{subfigure}{.185\textwidth}
    \includegraphics[width=\linewidth]{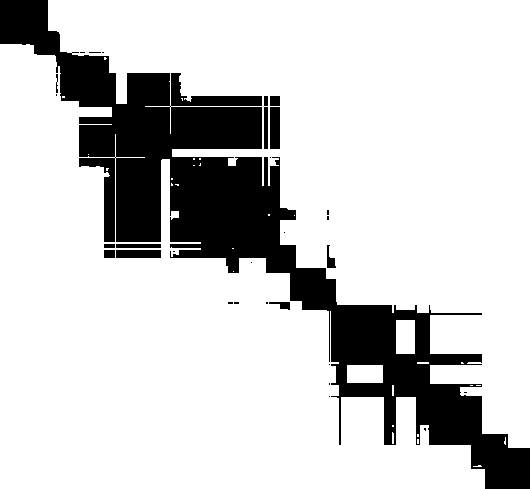}
    \caption{$I(\chain, \eps_3)$}
    \label{fig:matrix3}
\end{subfigure}
\begin{subfigure}{.185\textwidth}
    \includegraphics[width=\linewidth]{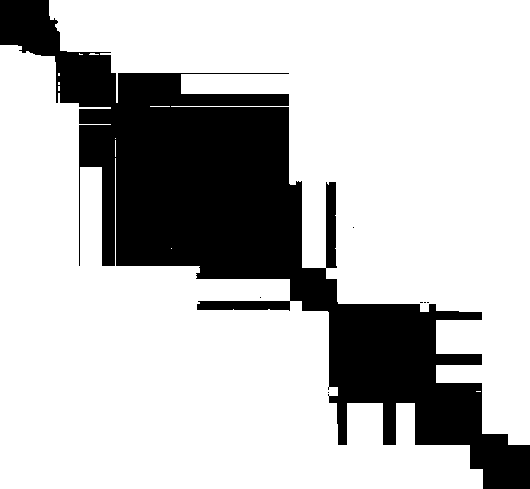}
    \caption{$I(\chain, \eps_4)$}
    \label{fig:matrix4}
\end{subfigure}
\begin{subfigure}{.185\textwidth}
    \includegraphics[width=\linewidth]{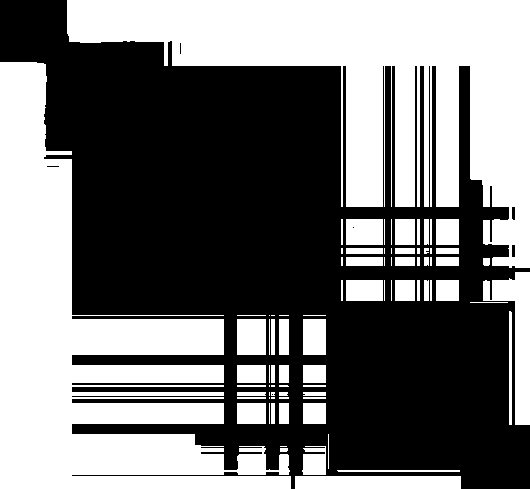}
    \caption{$I(\chain, \eps_5)$}
    \label{fig:matrix5}
\end{subfigure}
\caption{Shortcut intervals sets of a movement trajectory for five different error tolerance values. Each black cell represents a shortcut.}
\label{fig:shortcut_intervals_matrix}
\end{figure*}

To illustrate why this representation is useful, consider Figure~\ref{fig:shortcut_intervals_matrix}. Here, we see a shortcut interval set given by a movement trajectory for several different error tolerance values. The shortcut interval set is represented as a matrix, where the shading of a cell $(i, j)$ with $1\le i,j \le n$ indicates whether $(p_i, p_j)$ is a valid shortcut. Observe that regardless of the error tolerance, every column or row within the matrix only has a few shaded regions. We therefore can expect any $I_i(\eps)$ to be of constant size in practice, meaning $I(\chain, \eps)$ has linear space complexity in experimental settings. Shortcut interval sets are thus typically an order of magnitude smaller as opposed to storing the shortcut graph explicitly.

\begin{figure}[h!tb]
	\includegraphics[width=\linewidth]{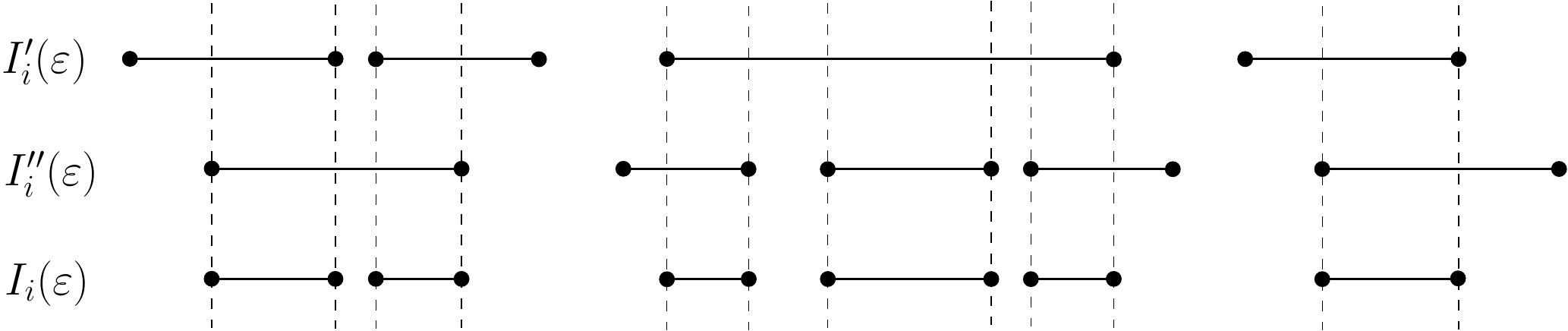}
	\caption{Overlapping two shortcut interval sets for some point $p_i$.}
	\label{fig:shortcut_intervals_intersection}
\end{figure}

For the Hausdorff distance, we can easily adapt the algorithm by Chan and Chin to efficiently construct shortcut intervals (see Figure \ref{fig:shortcut_intervals_intersection}).
Chan~and~Chin~\cite{chan1996approximation} proposed an algorithm for efficiently computing shortcut graphs under the Hausdorff distance in $O(n^2)$ time using a fixed error tolerance. This algorithm first computes two sets of shortcuts, and then intersects these sets to obtain the shortcut graph. Because both sets have a size of $O(n^2)$, this intersection runs in $O(n^2)$ time.

We can speed up the intersection of these sets by using shortcut intervals instead of representing the shortcuts explicitly. More specifically, we have two shortcut interval sets $I'(\chain, \eps)$ and $I''(\chain, \eps)$, and we wish to obtain an interval set $I(\chain, \eps)$ such that each $I_i(\eps)$ contains the overlap of the intervals in $I'_{i}(\eps)$ with the intervals of $I''_{i}(\eps)$. We can efficiently do this by simultaneously stepping through the sequence of shortcut intervals of $I'_i(\eps)$ and $I''_i(\eps)$, while computing every overlap encountered in $O(1)$ time. Because each set typically contains $O(n)$ shortcut intervals, this  then takes $O(n)$ time. An example is shown in Figure~\ref{fig:shortcut_intervals_intersection}.

The last step in computing a simplification for a given error is to compute a shortest path from $p_1$ to $p_n$ in the shortcut graph. This is typically done using breadth-first search in $O(n^2)$ time in practice. We show now how we can improve this by using the typically linear complexity of shortcut interval sets to find shortest paths in (typically) $O(n \log n)$ time in practice.

Consider we wish to find a shortest path in shortcut graph $G(\chain, \eps)$ from $p_s$ to $p_t$. We construct a balanced binary search tree $T$ containing all points $\langle p_s, \ldots, p_t \rangle$ ordered by their indices. For any point $p_r \in T$ rooted at subtree $T_r$, our objective is to annotate $p_r$ with the next point in a shortest path from $p_r$ to $p_t$ (and the length of this path), and the first point in a shortest path from any point in $T_r$ to $p_t$ (and the length of this path). Hence, there exists an annotation for every node and subtree in the binary search tree $T$.

\begin{figure*}[h!tb]
	\centering
	\includegraphics[width=\textwidth]{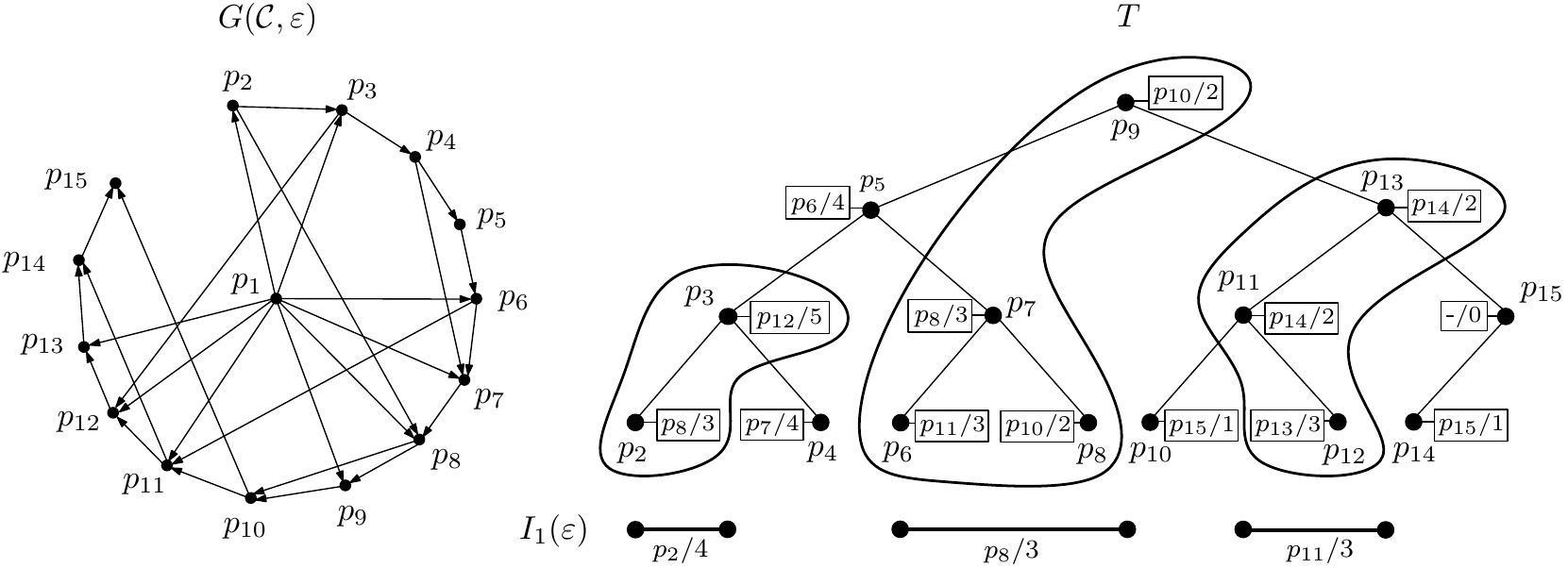}
	\caption{Finding the shortest path from $p_1$ to $p_{15}$ by means of range queries. We denote each node annotation as $p_x / \ell$, where $p_x$ is the next step in a shortest path from $p_1$ to $p_{15}$ with a length of $\ell$. Below every interval in $I_1(\eps)$ we show the candidate node annotation of $p_1$ given by the range query on that interval. In this case, $p_1$ is given a node annotation of either $p_8$ or $p_{11}$.}
	\label{fig:intervals_sssp}
\end{figure*}

We achieve a complete annotation of $T$ by inserting all points from $p_t$ down to $p_s$. Before inserting a point $p_i$, we perform a range query on $T$ for every shortcut interval in $I_i(\varepsilon)$. A range query for a shortcut interval $[x, y] \in I_i(\eps)$ finds all subtrees $\langle p_a, \ldots, p_b \rangle$ where $x \leq a \leq b \leq y$. For each point $p_j$ in each of these subtrees, there exists a shortcut $(p_i, p_j)$; we therefore can use the subtree's annotation to obtain a shortest path candidate from $p_i$ to $p_t$. After computing all path candidates, we insert $p_i$ and annotate it with the shortest path found, and maintain a valid tree annotation by updating the subtree annotation of all ancestors of $p_i$.   An example is provided in Figure~\ref{fig:intervals_sssp}.
Finally, after inserting all points, we construct the shortest path from $p_s$ to $p_t$ using the node annotation of $p_s$.

As aforementioned mentioned, we assume that $|I_i(\eps)|$ has constant size in practice for any point $p_i \in \chain$. However, it may occur that certain shortcut intervals are so small that performing the corresponding range query is more time-consuming than simply checking the node annotation of all points in that interval. There are $O(n^2)$ shortcut intervals in the worst case, yielding the worst-case running time of $O(n^2 \log n)$, which is slower than breadth-first search. Therefore,  we employ an optimisation for shortcut intervals $[x, y]$ where \mbox{$y - x < c \cdot \log n$} for some positive constant $c$ where we compute the shortest path by brute force in $O(y - x)$ time. By doing so, we reduce the worst-case running time of this algorithm to $O(n^2)$.

\section{Experimental Evaluation}\label{sec:experiments}
Varying spatial scales is central to many applications in the analysis of movement data.
One motivation for studying progressive simplification is to interactively explore trajectories at multiple levels of detail. For example, popular map services\footnote{\url{https://developers.google.com/maps/documentation/maps-static/intro\#Zoomlevels}}\footnote{\url{https://wiki.openstreetmap.org/wiki/Zoom_levels}} use many but constant number of spatial scales.

Firstly, we are interested how multiple scales influence the computation of simplification algorithms.
We want to evaluate our progressive simplification algorithm and compare it to existing simplification algorithms. In our experiments, we exploit how the input length, the number of scales, the number of shortcuts, and the error tolerance values impact the running time and the simplification size (cumulative as well as at each scale). Furthermore, we investigate how the performance differs for computing simplifications top down (zooming in) versus bottom up (zooming out).
Secondly, we evaluate the efficiency and the trade-offs between employing our convex hull construction with shortcut intervals and the explicit construction of shortcut graphs.

We use a movement trajectory of a migrating griffon vulture~\cite{movebank} in all our experiments that is highly suitable for multi-scale simplification, due to its high granularity and large distance span. We conducted the experiments on a 64-bit Intel Core i7-2630QM machine with 8 gigabytes of DDR3 SDRAM. All code was written in C\# 6.0 and is available at \url{https://github.com/WimReddingius/MultiScaleTrajectories}.



\subsection{Progressive Simplification}
We start our analysis by comparing the various simplification algorithms in a progressive simplification setting. We implemented these algorithms (compare Section~\ref{sec:introduction}): the optimal min-$\#$ algorithm by Imai~and~Iri \cite{Imai88} \emph{(II)}, the Douglas-Peucker simplification~\cite{douglas1973algorithms}~\emph{(DP)}, and the heuristic by Cao~et~al.~\cite{cao2006spatio}.

We use a sample of 5000 points and 10 scales, for which the associated error tolerance values are linearly sampled from the 10\% smallest shortcut errors. This set-up allows us to find simplifications that resemble the original curve well and emulates a similar number of scales as in state-of-the-art map services. All shortcut graphs are constructed using Chan~and~Chin's algorithm~\cite{chan1996approximation} and represented as shortcut interval sets.

We compare the optimal algorithms with greedy heuristics for progressive simplification running in $O(n^2m)$ time. Our optimal algorithm
constructs progressive simplifications top to down~\emph{(TD)}, but its performance is limited by the time spent determining the weights of all edges in the shortcut graphs. This can be speed up by greedily simplifying every simplification $\simpl_k$ using the shortcut graph $G(\chain, \eps_k)$ and propagating this choice to lower scales, imposing $\simpl_k \sqsubseteq \simpl_{k - 1}$. Alternatively, we can construct a progressive simplification from the bottom up~\emph{(BU)} by ensuring that all vertices in $\simpl_k$ are also present in $\simpl_{k - 1}$. We achieve this by skipping all shortcuts $(p_i, p_j)$ during construction of $G(\chain, \eps_k)$ where $p_i \not \in \simpl_{k - 1}$ or $p_j \not \in \simpl_{k - 1}$. Cao~et~al.~\cite{cao2006spatio} proposed an alternative bottom-up heuristic which uses any algorithm to produce $\simpl_k$, and use this simplification as input for the next round which constructs $\simpl_{k + 1}$. By doing so, $\simpl_{k+1} \sqsubseteq \simpl_{k}$ is imposed without doing any additional work. Note how this heuristic is different from the aforementioned bottom-up strategy, as it computes simplification $\simpl_k$ using the graph $G(\simpl_{k - 1}, \eps_k)$ instead of $G(\chain, \eps_k)$. Although faster, this heuristic runs the risk of progressively increasing the error of these simplifications with respect to the input curve. Specifically, the error of $\simpl_{k}$ may be in the worst case $\sum_{\ell = 1}^k \eps_\ell$ instead of just $\eps_k$.

Our implementation of the greedy algorithms use range queries on these shortcut interval sets to find shortest paths (implemented using left-leaning red black trees~\cite{leftleaning}), whereas the minimal simplification algorithm (Section~\ref{sec:progr_simplification}) uses Dijkstra's algorithm implemented using pairing heaps for priority queues~\cite{pairing}.

\begin{figure}[htb]
 \centering
 \includegraphics[width=0.8\linewidth]{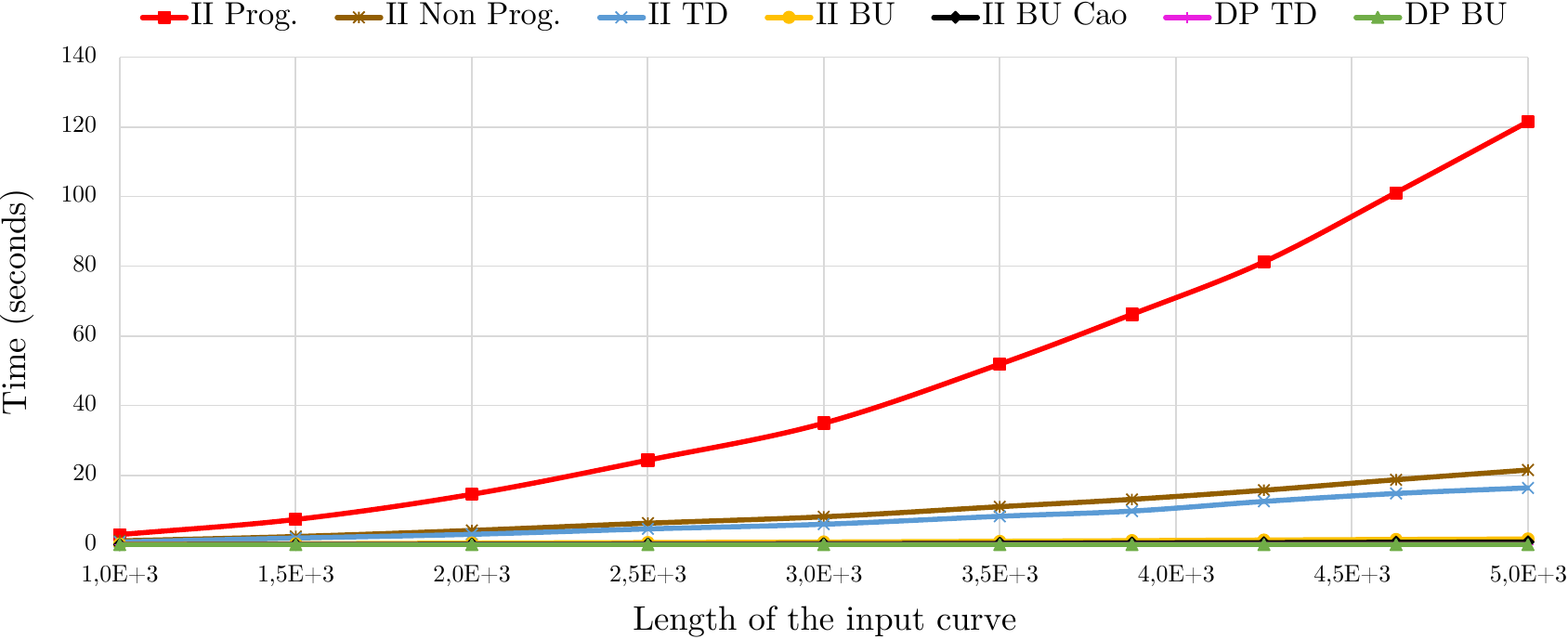}
 \caption{Running time in seconds with respect to the length of the input curve.}
 \label{fig:progr_time}
\end{figure}

\begin{figure}[htb]
 \centering
 \includegraphics[width=0.8\linewidth]{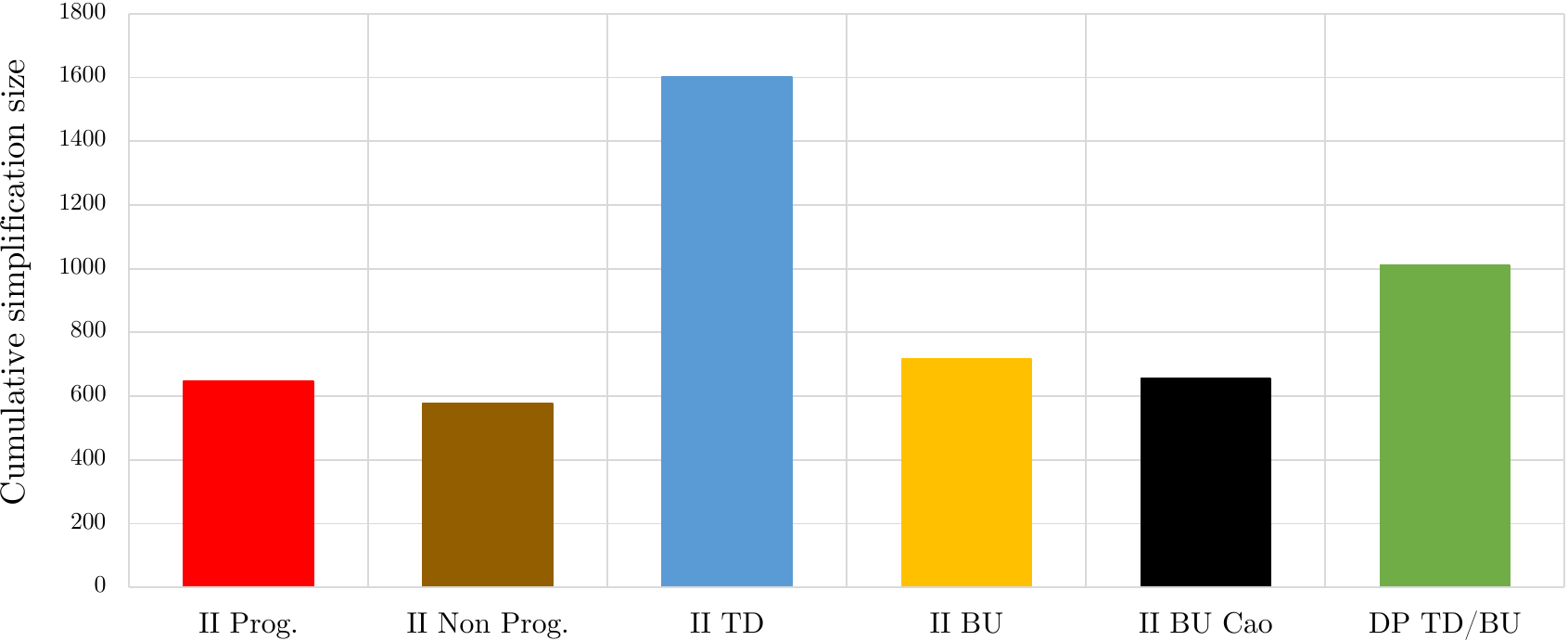}
 \caption{Cumulative simplification size for 5000 points.}
  \label{fig:hierarchy_amount_cumulative}
\end{figure}

In Figure~\ref{fig:progr_time} and Figure~\ref{fig:hierarchy_amount_cumulative}, we can see the running time and the cumulative simplification size for each simplification algorithm. As expected, the minimal progressive simplification algorithm \emph{(II Prog.)} is close in size to a minimal non-progressive simplification \emph{(II Non Prog.)}, but is at least an order of magnitude slower than the other algorithms. In Figure~\ref{fig:progr_time}, some lines overlay in the plot due to similar running times.

\begin{figure}[h!tb]
	\centering
	\includegraphics[width=0.8\linewidth]{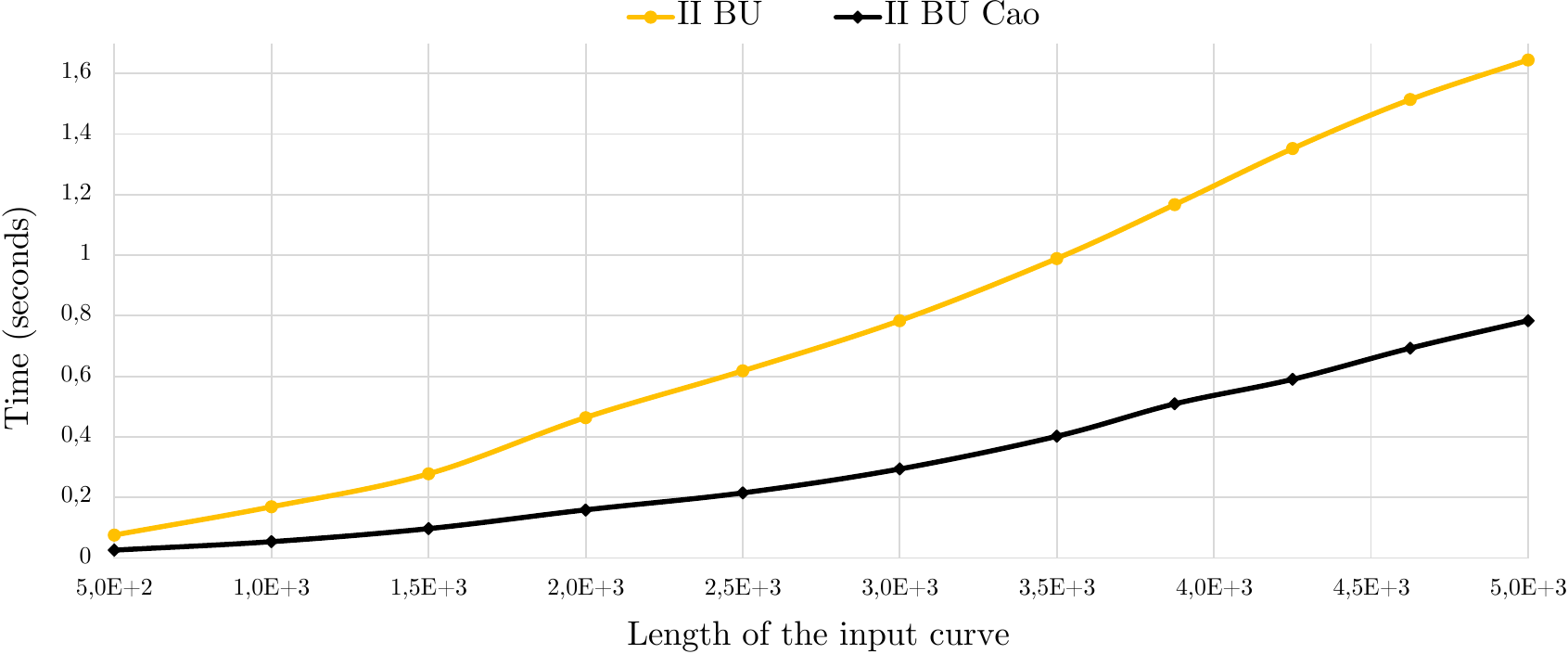}
	\caption{Running time of II BU and II BU Cao with respect to the length of the input curve.}
	\label{fig:greedy_iibu}
\end{figure}

Note that greedily constructing progressive simplifications from top to down \emph{(II TD)} yields a simplification size that is significantly larger than any other algorithm. This is due to inaccuracies of greedy choices at higher (coarser) scales that propagate to the simplifications on lower (finer) scales. A bottom-up construction \emph{(II~BU)} yields better results by starting with the least aggressive greedy choice at the lowest scale. \emph{II~BU} is also better than \emph{II~TD} in terms of running time, because the shortcut graphs for smaller error tolerance values are faster to construct, while greedy choices at these lower scales allow for drastic pruning during construction of the shortcut graphs at higher scales.
Our implementation of the bottom-up approach by Cao \emph{(II~BU~Cao)} results in marginally smaller simplifications than \emph{II~BU}, and ran 2-3$\times$ faster than \emph{II~BU} (see Figure~\ref{fig:greedy_iibu} for a plot), at the cost of a larger simplification error at higher scales.

\begin{figure}[h!tb]
	\centering
	\includegraphics[width=0.8\linewidth]{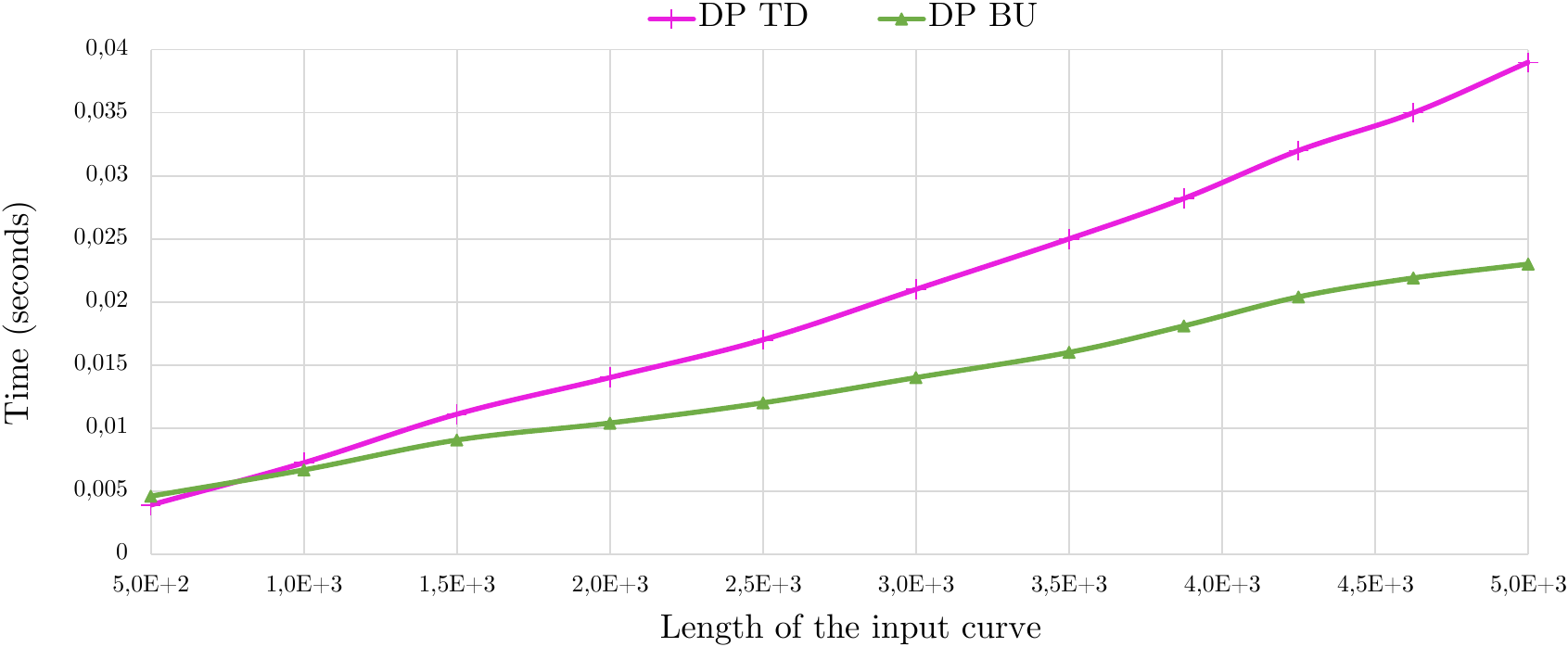}
	\caption{Running time of DP TD and DP BU with respect to the length of the input curve.}
	\label{fig:greedy_dp}
\end{figure}

We now evaluate how these approaches employing shortcut graphs perform against the efficient heuristic-based algorithm by Douglas and Peucker~\cite{douglas1973algorithms}. We substitute the simplification routine in \emph{II~TD} by \emph{DP~TD} and we do the same by replacing \emph{II~BU~Cao} with \emph{DP~BU}. Because \emph{DP~BU} uses the progressive simplification heuristic by Cao~et~al.~\cite{cao2006spatio}, we would expect the simplification error to become more severe at higher scales, much like \emph{II~BU~Cao}. However, because Douglas-Peucker simplification recursively splits the input curve at specific points consistently until the given error is reached, this does not occur. It is also this splitting strategy that always yields \emph{DP~TD} and \emph{DP~BU} the same exact progressive simplification. Because Douglas-Peucker simplification has near-linear performance in practice, \emph{DP~TD} and \emph{DP~BU} are also exceptionally fast, with running times around 40 times lower than \emph{II~BU}, wherein \emph{DP~BU} is significantly faster than \emph{DP~TD} (see Figure~\ref{fig:greedy_dp}). However, because the simplification size is minimized heuristically, these algorithms produce larger simplifications when compared to using shortcut graphs (with the exception of \emph{II~TD}).

\begin{figure}[htb]
 \centering
 \includegraphics[width=0.8\linewidth]{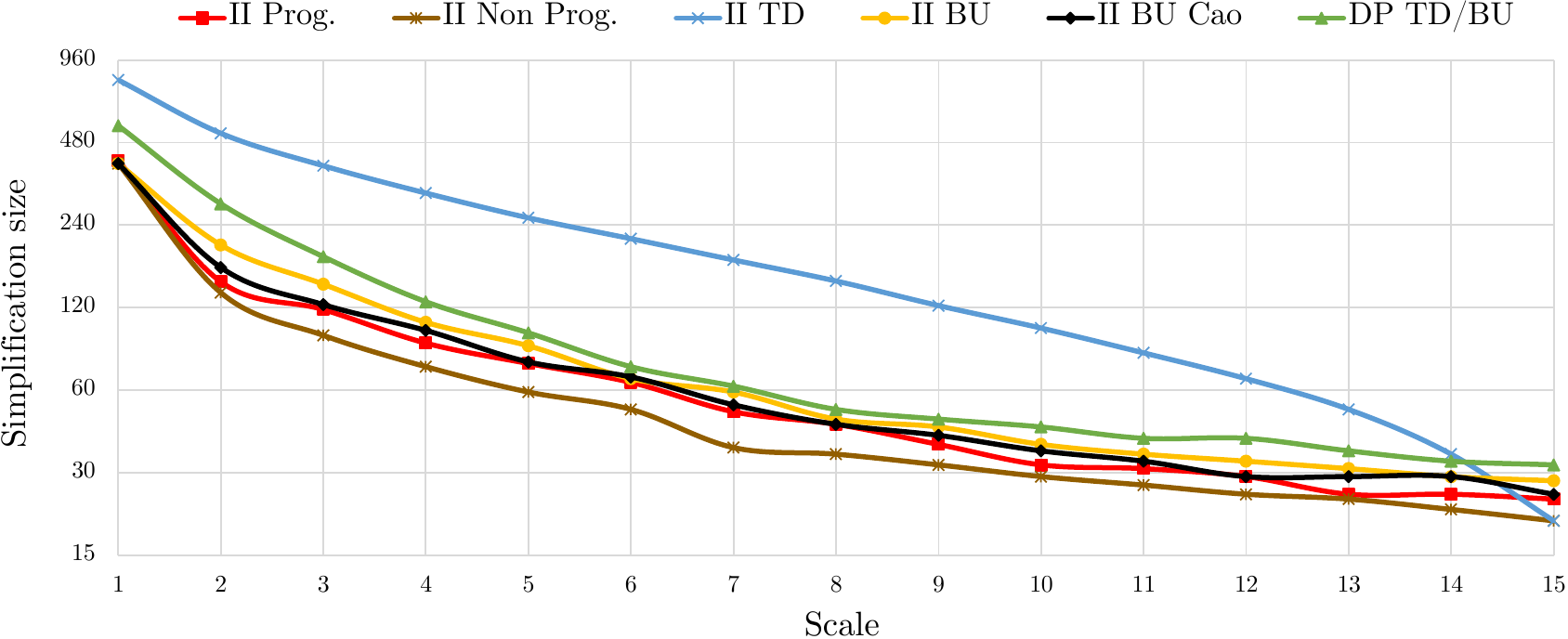}
 \caption{Simplification size at every scale on a logarithmic scale.}
 \label{fig:size_by_scale}
\end{figure}

In Figure~\ref{fig:size_by_scale}, we gain insights in
how the cumulative simplification size is distributed over all scales for each algorithm. \emph{II~TD} is of particular interest, because we can see just how fast the greedy choices at higher scales propagate to cause inaccuracies at lower scales. This results in larger simplification sizes than any other algorithm. \emph{II~TD} reaches the smallest simplification at scale~15, yet the previous greedy choices at scales 1 to 14 yield the largest cumulative simplification size of all algorithms (compare Figure~\ref{fig:hierarchy_amount_cumulative}).


\subsection{Shortcut graphs}
We now conduct experiments on techniques to construct shortcut graphs at multiple scales.
For this, we compare the running time of constructing the shortcut graphs independently using Chan and Chin's algorithm~\cite{chan1996approximation} with shortcut interval sets, and integrated construction using convex hulls (Section \ref{sec:convex_hulls}). Left-leaning red-black trees~\cite{leftleaning} are used for representing the convex hull. We use an input curve of length 3500, and the errors $\eps_k$ are chosen by linearly sampling from \emph{all} shortcut errors. A linear sampling allows us to investigate the link between the errors and the number of shortcuts. While the independent construction requires 7-8 seconds per shortcut graph, the integrated construction requires around 280 seconds for pre-computing all errors. In our experiments, this pre-computation starts paying off at around 65 scales, making it worthwhile for constructing minimal continuous progressive simplifications, which may require a quadratic number of scales.

Next, we evaluate shortcut intervals, i.e., our approach to compress shortcut graphs (cf.~Section~\ref{sec:intervals}), which may speed up both progressive and non-progressive  simplification algorithms that use the shortcut graph. We analyze the space complexity, the construction time and the running time of shortest path calculations. We perform experiments along two dimensions: the length of the input curve and the simplification error. We investigate the latter, since the level of compression that can be obtained by using shortcut intervals highly depends on the density of the shortcut graph, which is directly related to the used error tolerance. Shortcut graphs with higher density typically need less shortcut intervals. This phenomenon can be observed in Figure~\ref{fig:shortcut_intervals_matrix}.

\begin{figure}[h!bt]
\centering
\includegraphics[width=0.8\linewidth]{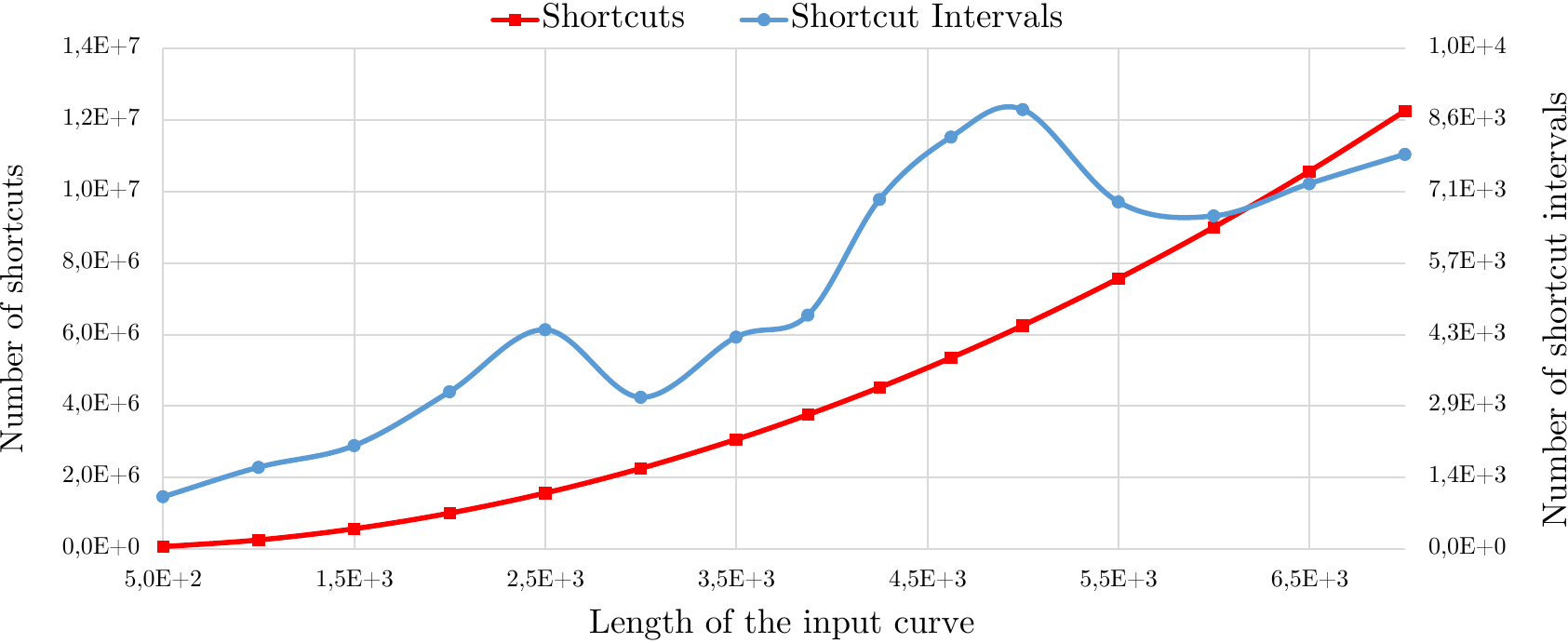}
 \caption{The space complexity of representing a shortcut graph with a density of 50\% by either explicitly storing all shortcuts, or using shortcut intervals, for various lengths of the input curve.}
 \label{fig:intervals_amount_by_n}
\end{figure}%

We start by analyzing how the input length influences the graph construction and the shortest path computation. First, we compare the space complexity of shortcut interval sets and explicit shortcut graphs. For this, we fix the density of the shortcut graph to 50\%. Figure~\ref{fig:intervals_amount_by_n} reveals that the number of shortcuts grows quadratically in the length of the input curve, whereas the number of shortcut intervals seems to grow linearly, though non-monotonically. This non-monotonic growth reflects the fact that the progression of the input curves changes as we extend the input sample.

\begin{figure}[h!tb]
	\centering
	\includegraphics[width=0.8\linewidth]{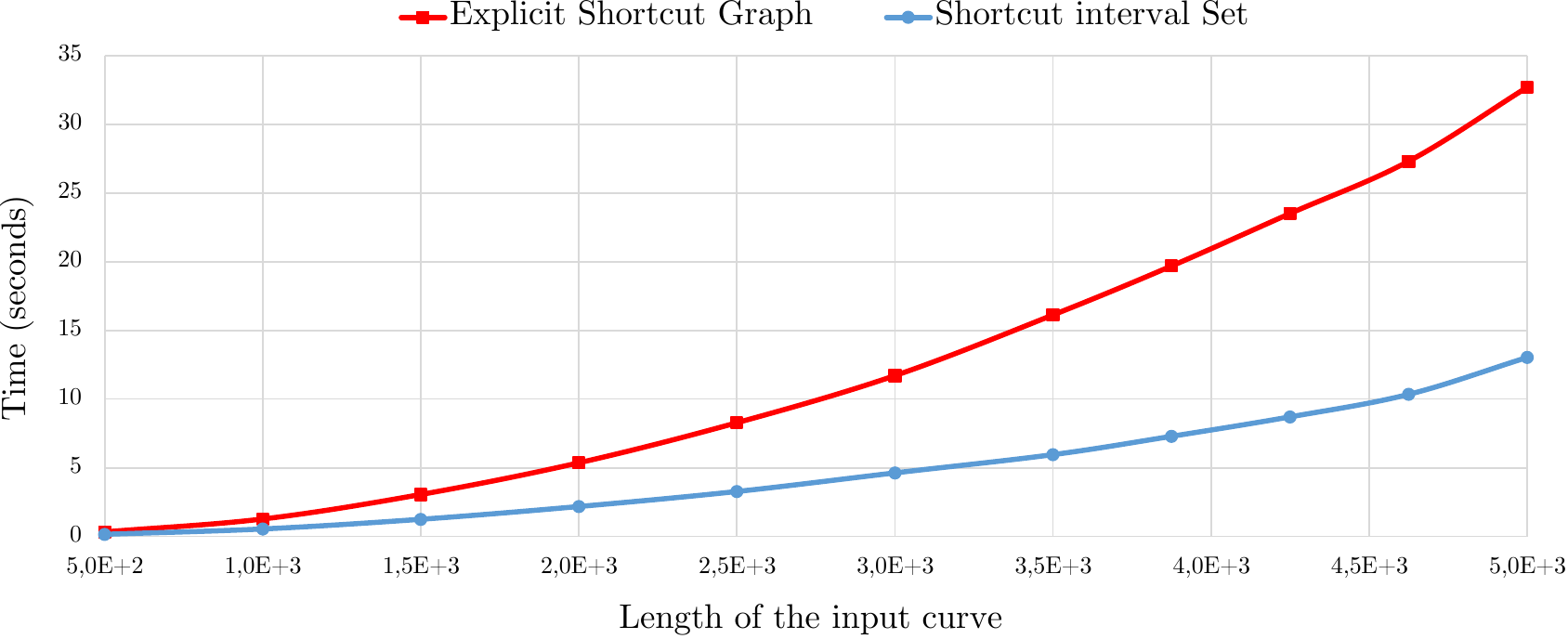}
	\caption{Running time of constructing a shortcut interval set or an explicit shortcut graph using Chan and Chin's algorithm~\cite{chan1996approximation} for various lengths of the input curve.}
	\label{fig:chin_chan_time_n}
\end{figure}

The time required to construct a shortcut graph under the Hausdorff distance using the algorithm by Chan and Chin turns out to be consistently more than twice as fast when using shortcut intervals instead of an explicit construction. These experiments were performed with a graph density of 25\% (see plot in Figure~\ref{fig:chin_chan_time_n}).

\begin{figure}[h!tb]
	\centering
	\includegraphics[width=0.8\linewidth]{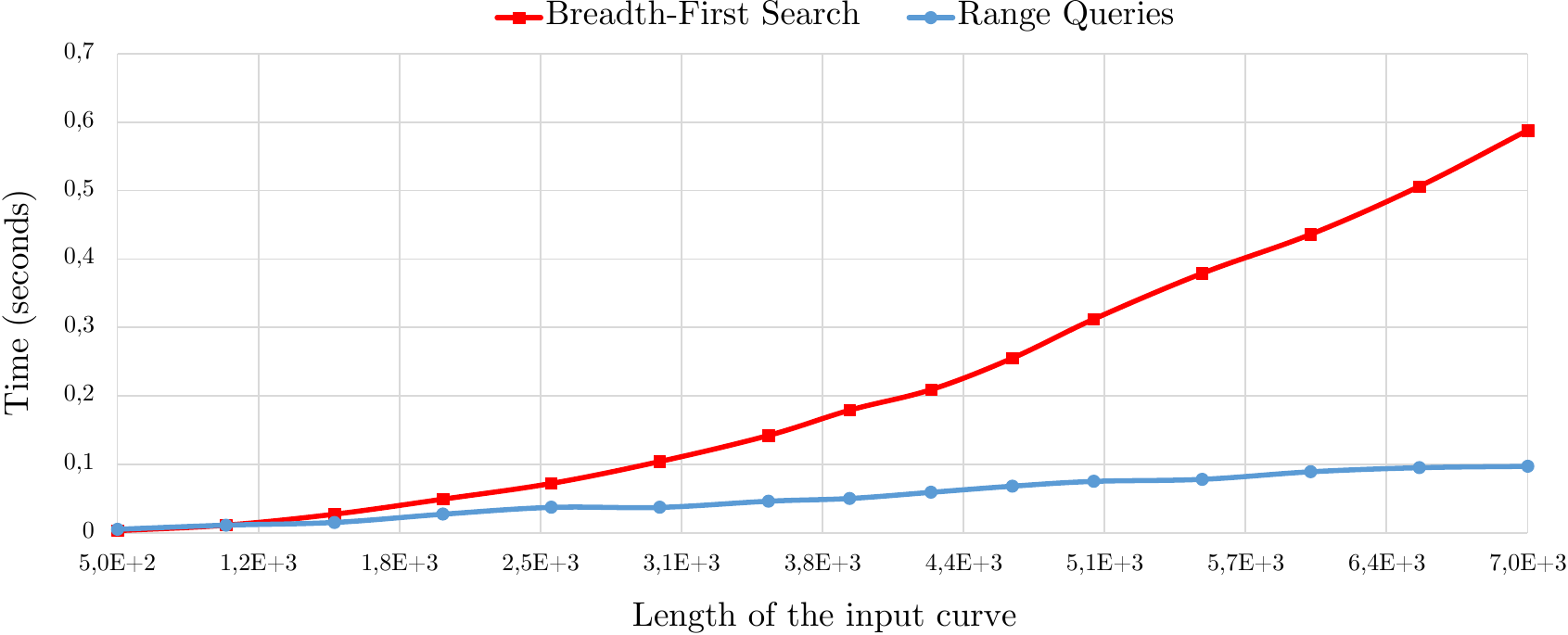}
	\caption{Running time for various lengths of the input curve of finding a shortest path from $p_1$ to $p_n$ in a shortcut graph with 50\% density using breadth-first search, or range queries on shortcut interval sets.}
	\label{fig:intervals_time_by_n}
\end{figure}

For finding shortest paths in shortcut graphs, recall that by using range queries on shortcut intervals, we spend time with respect to the number of shortcut intervals, whereas a breadth-first search affects the number of shortcuts. Figure~\ref{fig:intervals_time_by_n} shows that this relation also holds true in experiments. We observe that by using range queries on shortcut intervals, we spend near-linear time to find a shortest path in the shortcut graph. We foresee this improvement to be an important stepping stone towards computing (non-progressive) simplifications in near-linear time on large data.

\begin{figure}[htb]
	\centering
	\includegraphics[width=0.8\linewidth]{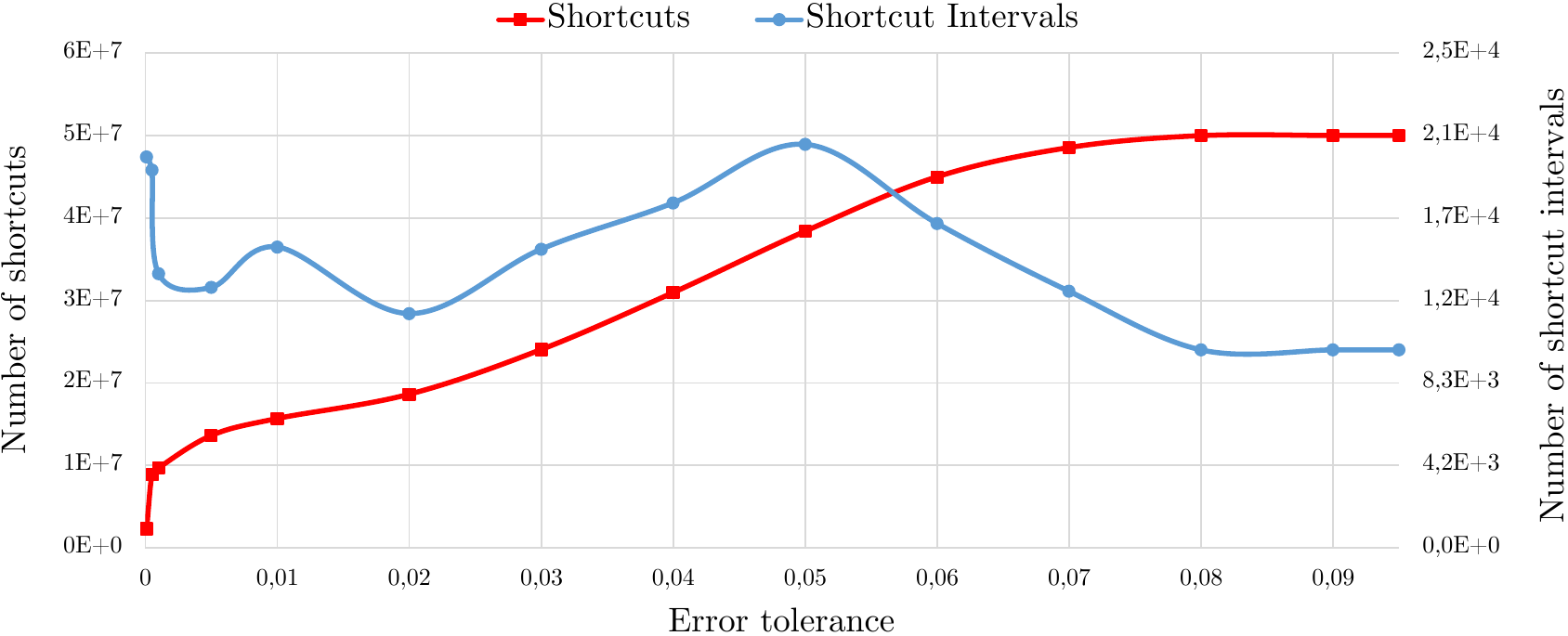}
	\caption{The space complexity of representing a shortcut graph for an input curve with 10\,000 points by either explicitly storing all shortcuts, or using shortcut intervals, for various error tolerances. The peek at $\eps = 0.05$ corresponds to a graph density of 80\%}
	\label{fig:intervals_amount_by_error}
\end{figure}%

\begin{figure}[htb]
	\centering
	\includegraphics[width=0.8\linewidth]{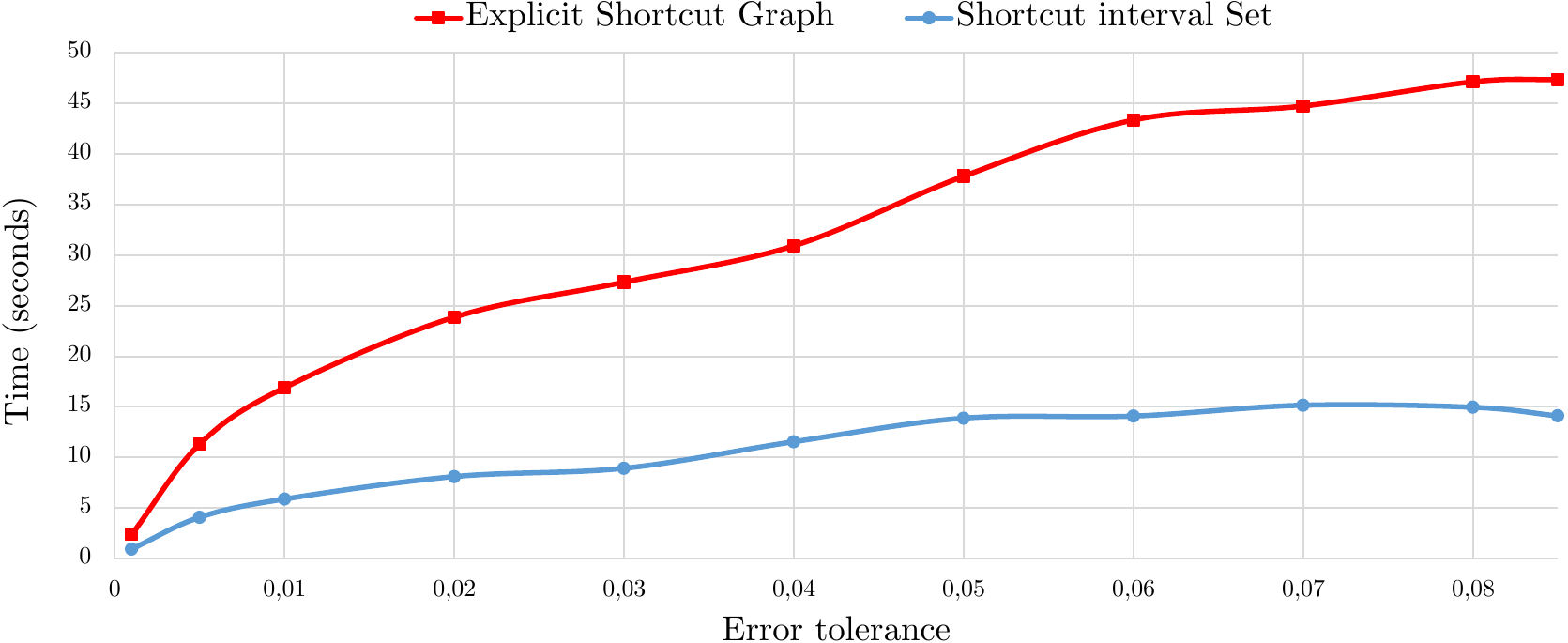}
	\caption{Running time of constructing a shortcut interval set or an explicit shortcut graph using Chan and Chin's algorithm~\cite{chan1996approximation} on an input curve with 3000 points for varying error tolerances.}
	\label{fig:chin_chan_time_error}
\end{figure}

\begin{figure}[h!tb]
	\centering
	\includegraphics[width=0.8\linewidth]{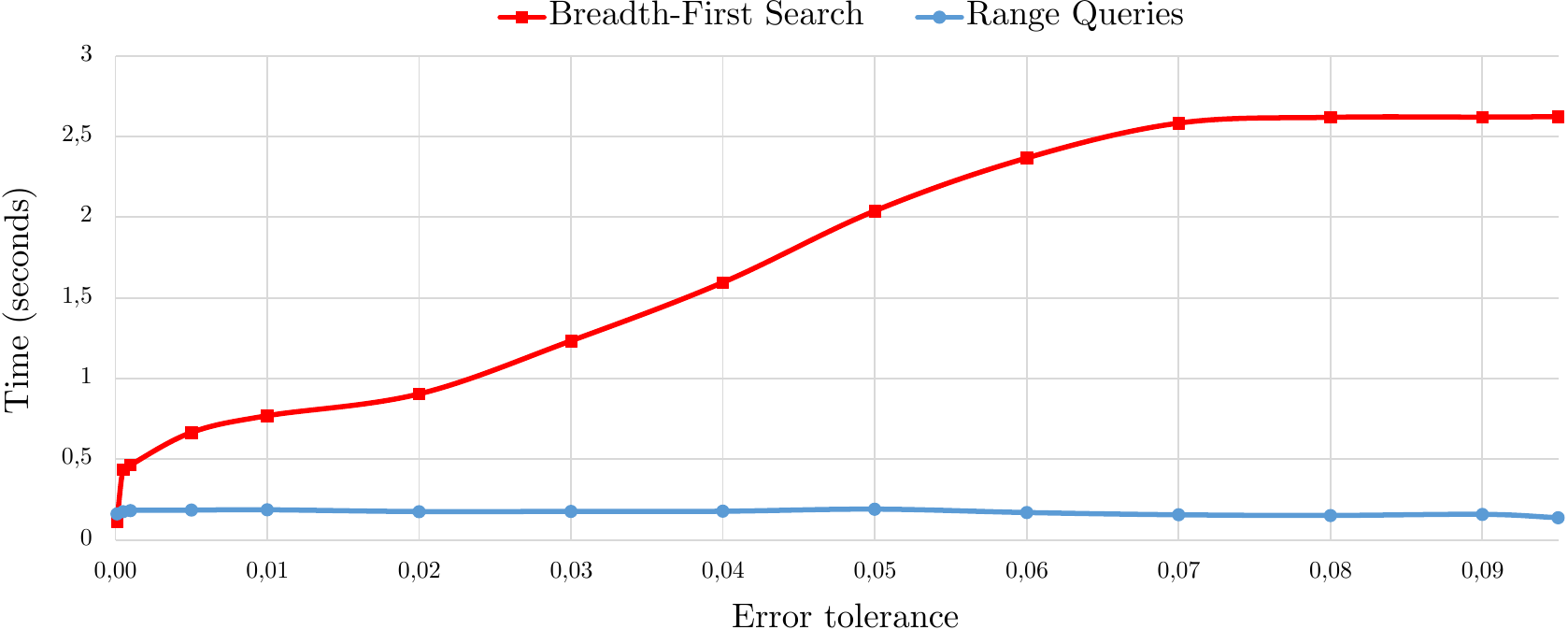}
	\caption{Running time for various error tolerances of finding a shortest path in a shortcut graph from $p_1$ to $p_n$ using breadth-first search or range queries on shortcut interval sets, on an input curve with 10\,000 points.}
	\label{fig:intervals_time_by_error}
\end{figure}

Next, we evaluate how the magnitude of the error tolerance influences these results (corresponding plots in Figures~\ref{fig:intervals_amount_by_error}, \ref{fig:chin_chan_time_error}, and \ref{fig:intervals_time_by_error}). The number of shortcut intervals increases and decreases in no discernible pattern, as the error tolerance grows. As discussed earlier, this is related to the growth in coarseness among the shortcut intervals as the shortcut graph density increases. We observe the monotonic growth in the number of shortcuts (see Figure~\ref{fig:intervals_amount_by_error}), whereas the number of shortcut intervals peeks around $\eps = 0.05$, which corresponds to a shortcut graph density of 80\%.

The construction of shortcut graphs using Chan and Chin's algorithm~\cite{chan1996approximation} is around three times as faster when using shortcut interval sets (see Figure~\ref{fig:chin_chan_time_error}), regardless of the error tolerance.
We suspect that the implementation of shortcut interval sets is inherently more efficient, since no index of shortcuts needs to be maintained to facilitate the intersection of shortcut sets.


The running time of finding shortest path using breadth-first search and range queries on shortcut interval sets is as expected directly related to the number of shortcuts and number of shortcut intervals respectively. Here, shortcut intervals show their strength since the number of shortcuts is typically an order of magnitude higher than the number of shortcut intervals for most choices for the error tolerance (see Figure~\ref{fig:intervals_time_by_error}).

\section{Conclusions}
We present the first algorithm for computing a progressive simplification with minimal complexity. For an input curve of $n$ vertices, this algorithm runs in $O(n^3m)$ time for $m$ discrete scales, and $O(n^5)$ time for continuous scaling.
Furthermore, we show how to compute the errors $\eps(p_i,p_j)$ for all shortcuts in $O(n^2 \log n)$ time under the Hausdorff distance.


The experimental evaluation on trajectory data shows that our progressive algorithm is effective, yet too slow for larger data, and provides similar cumulative simplification sizes as an optimal non-progressive simplification algorithm.
 Greedy construction of the progressive simplification from the bottom up is shown to provide a reasonable, faster alternative. Our experimental results further indicate that integrated construction of multiple shortcut graphs is effective when employed for many scales, and thus particularly useful for continuous progressive simplification. Finally, shortcut intervals show a significant reduction in memory usage, and allows for finding shortest paths in near-linear time in practice. However, applications use a constant number of scales up to now.

As future work, it would be of interest to improve the running time of the minimal progressive simplification algorithm to facilitate its application on large data. With the improvements made to finding shortest paths in the shortcut graph, we are one step closer to an algorithm that computes minimal (non-progressive) simplifications that also in running time is competitive with fast heuristics~\cite{douglas1973algorithms} and approximation algorithms~\cite{AgarwalHMS05}. To realize this, we need new and efficient techniques for constructing shortcut interval sets, in particular for large error tolerance. Finally, it would be interesting to see whether our global optimization approach can be extended and is effective for progressively meshing surfaces~\cite{hoppe1996progressive}.

{\small
\subparagraph*{Acknowledgements.}
We thank Michael Horton for our discussions on this topic.
}

\bibliography{references}

\end{document}